\documentclass[onecolumn]{mnras}

\usepackage{newtxtext,newtxmath}

\usepackage[T1]{fontenc}

\DeclareRobustCommand{\VAN}[3]{#2}
\let\VANthebibliography\thebibliography
\def\thebibliography{\DeclareRobustCommand{\VAN}[3]{##3}\VANthebibliography}

\usepackage{graphicx}	
\usepackage{amsmath}	

\usepackage{ulem}
\usepackage{times}
\usepackage[usenames,dvipsnames]{pstricks}
\usepackage{psfrag}
\usepackage{lipsum}

\usepackage{textcase}

\usepackage{booktabs}
\usepackage{comment}
\usepackage{textgreek}
\usepackage{hyperref}

\usepackage{apalike}
\let\bibhang\relax
\usepackage{natbib}

\usepackage{wrapfig}
\usepackage{lscape}
\usepackage{rotating}

\def\araa{ARA\&A}
\def\apj{ApJ}
\def\apjl{ApJ}
\def\apjs{ApJS}
\def\aap{A\&A}
\def\mnras{MNRAS}
\def\pasj{PASJ}
\def\nat{Nature}

\newcommand{\be}{\begin{equation}}
\newcommand{\ee}{\end{equation}}
\newcommand{\bary}{\begin{eqnarray}}
\newcommand{\eary}{\end{eqnarray}}

\title[Microphysical Parameters and Afterglows]{Microphysical Parameter Variation in GRB Stratified Afterglows and Closure Relations: from sub-GeV to TeV Observations}

\author[N. Fraija et al.]{
Nissim Fraija,$^{1}$\thanks{E-mail: nifraija@astro.unam.mx}
Maria G. Dainotti,$^{2,3,4}$
Boris Betancourt Kamenetskaia,$^{5,6}$
\newauthor Antonio Galv\'an-G\'amez,$^{1}$
Edilberto Aguilar-Ruiz$^{1}$
\\
$^{1}$Instituto de Astronom\' ia, Universidad Nacional Aut\'onoma de M\'exico,\\ Circuito Exterior, C.U., A. Postal 70-264, 04510 M\'exico City, M\'exico\\
$^{2}$National Astronomical Observatory of Japan, 2-21-1 Osawa, Mitaka, Tokyo 181-8588, Japan\\
$^{3}$Space Science Institute, Boulder, CO, USA\\
$^{4}$The Graduate University for Advanced Studies, SOKENDAI, Shonankokusaimura, Hayama, Miura District, Kanagawa 240-0193, Japan\\
$^{5}$Technical University of Munich, TUM School of Natural Sciences, Physics Department, James-Franck-Str 1, 85748 Garching, Germany\\
$^{6}$Max-Planck-Institut f\"ur Physik (Werner-Heisenberg-Institut), F\"ohringer Ring 6, 80805 Munich, Germany
}

\date{Accepted XXX. Received YYY; in original form ZZZ}

\pubyear{2015}

\begin{document}
\label{firstpage}
\pagerange{\pageref{firstpage}--\pageref{lastpage}}
\maketitle

\begin{abstract}
Gamma-ray bursts (GRBs) are one of the most exciting sources that offer valuable opportunities for investigating the evolution of energy fraction given to magnetic fields and particles through microphysical parameters during relativistic shocks.   The delayed onset of GeV-TeV radiation from bursts detected by the \textit{Fermi} Large Area Telescope (\textit{Fermi}-LAT) and Cherenkov Telescopes provide crucial information in favor of the external-shock model. Derivation of the closure relations (CRs) and the light curves in external shocks requires knowledge of GRB afterglow physics. In this manuscript, we derive the CRs and light curves in a stratified medium with variations of microphysical parameters of the synchrotron and SSC afterglow model radiated by an electron distribution with a hard and soft spectral index. Using Markov Chain Monte Carlo simulations,  we apply the current model to investigate the evolution of the spectral and temporal indexes of those GRBs reported in the Second Gamma-ray Burst Catalog (2FLGC), which comprises 29 bursts with photon energies above 10 GeV and of those bursts (GRB 180720B, 190114C, 190829A and 221009A) with energetic photons above 100 GeV, which can hardly be modeled with the CRs of the standard synchrotron scenario. The analysis shows that i) the most likely afterglow model using synchrotron and SSC emission on the 2FLGC corresponds to the constant-density scenario, and ii)  variations of spectral (temporal) index keeping the temporal (spectral) index constant could be associated with the evolution of microphysical parameters, as exhibited in GRB 190829A and GRB 221009A.
\end{abstract}





\section{Introduction}


Gamma-ray bursts (GRBs), the most extreme phenomena in the universe that emit non-repeating gamma-ray  flashes,\footnote{These flashes are usually reported in the keV-MeV energy range and modeled by the empirical Band function \citep{1993ApJ...413..281B}.} originate in the core collapse of massive stars \citep{1993ApJ...405..273W,1998ApJ...494L..45P, Woosley2006ARA&A} or the merger of binary compact objects, like a black hole (BH) - neutron star (NS) \citep{1992ApJ...395L..83N}, a BH - BH or a  NS - NS \citep{1992ApJ...392L...9D, 1992Natur.357..472U, 1994MNRAS.270..480T, 2011MNRAS.413.2031M}.  These scenarios can lead to a hyper-accreting BH or a millisecond magnetar\footnote{A fast-spinning newly born highly-magnetized NS.} as a remnant.  Their classification is determined by the duration of the main gamma-ray episode ($T_{90}$)\footnote{$T_{90}$ represents the period during which a GRB emits between $5\%$ and $95\%$ of the total observed counts from its prompt emission.} as short (sGRBs) and long GRBs (lGRBs). Based on the nature of progenitors, sGRBs are associated with the merger of binary compact objects and lGRBs to the core collapse of massive stars.  Irrespective of the progenitor, a late episode, known as afterglow, is observed from radio wavelengths to TeV gamma-rays. It is usually interpreted in the synchrotron afterglow scenario, generated by non-thermal electrons accelerated in the forward shock. This shock originates  when the outflow encounters the circumburst medium and transfers part of its energy to it \citep{1999ApJ...513..669K, 2002ApJ...568..820G, 1998ApJ...497L..17S}. Synchrotron photons are scattered up to energies greater than tens of GeV by the synchrotron self-Compton (SSC) mechanism \citep{2001ApJ...559..110Z, 2019ApJ...883..162F}, even if the highest photon energy produced by the synchrotron process is $\sim 5 - 10 ~{\rm GeV}$ \citep{2010ApJ...718L..63P, 2009ApJ...706L.138A, 2011MNRAS.412..522B}.

The Major Atmospheric Gamma Imaging Cherenkov \citep[MAGIC;][]{MAGIC_2019Natur.575..455M, 2019Natur.575..459A, 2021ApJ...908...90A, 2022icrc.confE.788F},  the High Energy Stereoscopic System \citep[H.E.S.S.;][]{2019Natur.575..464A, HESS_2021Sci...372.1081H} and the Large High Altitude Air Shower Observatory \citep[LHAASO;][]{2023Sci...380.1390L} reported very-high-energy (VHE) emission with energetic photons above 100 GeV in GRB 160821B, 180720B, 190114C, 190829A, 201216C and 221009A. \cite{2019ApJ...878...52A} has released the \textit{Fermi} Collaboration's Second Gamma-ray Burst Catalog (2FLGC), which contains data from the first decade of their program (from 2008 to 2018, August 4). The data set contains 169 GRBs with high-energy emission $\ge100$ MeV and a sample of GRBs with temporarily-extended emission fitted by a power-law (PL) and a broken power-law (BPL) function.  A BPL function with a temporal break at hundreds of seconds characterizes several light curves.  In most cases, the temporarily-extended emission is explained according to the variation in temporal and spectral indices called the closure relations (CRs) predicted by the synchrotron afterglow model. Although it is generally accepted that  CRs will be present if the forward shock dominates the high-energy emission, this is not the case for all LAT light curves.

\cite{2019ApJ...883..134T} methodically evaluated the CRs in a sample of 59 LAT-detected bursts using temporal and spectral indices. While they found that the conventional synchrotron forward-shock emission successfully explained the spectral ($\beta_{\rm LAT}$) and temporal ($\alpha_{\rm LAT}$) indices in most cases, they also found that a sizable minority of bursts defied this explanation.  They found that a significant fraction of GRBs satisfies the CRs when the synchrotron afterglow model: i) is in the slow-cooling regime ($\nu^{\rm syn}_{\rm m}<\nu_{\rm LAT}<\nu^{\rm syn}_{\rm c}$), ii) occurs in the constant-density medium, and iii) has an abnormally low value of $\varepsilon_B<10^{-7}$. The synchrotron spectral breaks $\nu^{\rm syn}_{\rm c}$ and $\nu^{\rm syn}_{\rm m}$ are the cooling and characteristic breaks, respectively.   \cite{2021ApJS..255...13D} investigated the PL segment in the X-ray and optical light curves with a flatter slope than the typical decline, the so-called ``plateau" phase. This phase is seen as being consistent with the energy injection scenario \citep{2005ApJ...635L.133B,  2005ApJ...630L.113K, 2006Sci...311.1127D, 2006ApJ...636L..29P, 2006MNRAS.370L..61P, 2005Sci...309.1833B, 2007ApJ...671.1903C, 2017MNRAS.464.4399D, 2019ApJ...872..118B}. They concluded that the most advantageous scenario was one in which the constant-density environment underwent slow cooling. \cite{2022ApJ...934..188F} estimated the CRs for a forward-shock model of the SSC that originates in the stellar-wind and interstellar medium. The authors demonstrated that the synchrotron scenario could not explain a sizable proportion of GRBs and that the constant-density medium is favored without energy injection. In contrast, the stellar-wind medium is favored with the presence of injection.    Several studies have employed the development of CRs in the interstellar medium ($k=0$) and the stellar-wind ($k=2$) to verify the development of the LAT light curve. Nonetheless, a circumburst medium with a density profile between ${\rm k=0}$ and ${\rm k=2}$ is consistent in several GRBs, as revealed by studies of the progenitor and modeling of multiwavelength data during the afterglow phase.   \cite{2013ApJ...776..120Y} looked at the dynamics of forward/reverse shocks as they traveled through a medium represented by the PL distribution. Based on their model, the density profile index of 19 GRBs ranged from $ 0.4 \leq {\rm k}\leq  1.4$, with a mean value of  ${\rm k\sim1}$. Similar results were found by analyzing 146 GRBs instead of only 13, as shown in \cite{2013ApJ...774...13L}. \cite{2008Sci...321..376K} derived a density-profile index with $k>2$  after looking at the possibility that a BH's accretion of a stellar envelope causes the plateau phase in X-ray light curves.  Multiwavelength observations of GRB 130427A, one of the most energetic bursts at redshift $z< 1$, were modeled by \cite{2013ApJ...779L...1K} using a synchrotron afterglow model in the adiabatic domain without energy injection for $p>2$. The authors concluded that the data most aligned with a density profile value somewhere between ${\rm k}=0$ and ${\rm k}=2$.

Variations in the values of microphysical parameters ($\epsilon_{\rm e}$ and $\epsilon_{\rm B}$)\footnote{The micropysical parameters $\epsilon_{\rm e}$ and $\epsilon_{\rm B}$ correspond to the fraction of energy given to accelerate electrons and amplify the magnetic field, respectively.} have also been considered during the afterglow phase \citep[e.g., see][]{2003ApJ...597..459Y, 2003MNRAS.346..905K, 2006A&A...458....7I,2006MNRAS.369..197F, 2006MNRAS.369.2059P, 2005PThPh.114.1317I, 2006MNRAS.370.1946G, 2020ApJ...905..112F}. Given the lack of knowledge about how energy is transferred from protons to electrons and magnetic fields in relativistic shocks, this is not unrealistic to assume. \cite{2005PThPh.114.1317I} proposed that the fraction of energy given to accelerate electrons could evolve during the early afterglow when the synchrotron radiation was in the fast-cooling regime.   Considering that the efficiency of conversion of the energy carried by the outflow into radiation depends on the microphysical parameters, \cite{2006MNRAS.370.1946G} suggested that the plateau phase could be interpreted by the variation of the microphysical parameters instead of the energy injection right after the prompt episode. \cite{2006MNRAS.369..197F} reported that energy injection and the evolution of microphysical parameters are consistent with the afterglow observations. To mimic the brightenings observed in some GRB afterglows, \cite{10.1111/j.1365-2966.2009.15886.x} looked into the evolution of the microphysical parameters under the setting of a wind bubble environment. \cite{2020ApJ...905..112F} considered the variation of microphysical parameters to explain both features simultaneously, the rapid evolution of the synchrotron cooling break, and the X-ray plateau phase reported in GRB 160509A.   Given the failure of an early effort to implement inverse Compton cooling, \cite{2006MNRAS.369..197F} argued that it would be worthwhile to search for as-of-yet-unexamined mechanisms.

We derive the SSC and synchrotron forward-shock model including the corresponding light curves, spectra and CRs with the evolution of the microphysical parameters in the stratified environment with a density profile $\propto r^{-k}$ with ${\rm k}$ in the range of  $0\leq k < 3$. We apply the current model to investigate the evolution of the spectral and temporal indexes of the GRBs reported in 2FLGC, and of the bursts that emitted VHE emission.  We consider an electron distribution with a hard spectral index ($1\leq p \leq 2$) and $ 2 < p$. This manuscript is organized as follows: in Section \S\ref{sec2}, we derive the SSC forward-shock model with the evolution of the microphysical parameters in a stratified environment with $0\leq k < 3$. In Section \S\ref{sec3}, we introduce \textit{Fermi}-LAT data and the methodical approach with a discussion about  the variation of magnetic parameter in decaying microturbulence,  the circumburst environment,  the maximum synchrotron photons and the magnetic microphysical parameter.  In addition,  we derive and plot the expected light curves and spectral energy distributions (SEDs) and apply the theoretical model to describe the latest LAT observations of GRB 160509A.  In Section \S\ref{sec4}, we introduce and analyze the CRs of SSC afterglow model with the VHE observations of  GRB 180720B, 190114C, 190829A and 221009A.   In Section \S\ref{sec5}, we summarize previous works and point out the differences with current manuscript.   Finally, in Section \S\ref{sec5}, we conclude by summarizing our findings and making some general observations. From this point on, we will denote quantities as ``unprimed" and ``primed" if they are measured in the observer and the comoving frame, respectively.   We assume for the cosmological constants a spatially flat universe $\Lambda$CDM model with  $H_0=69.6\,{\rm km\,s^{-1}\,Mpc^{-1}}$, $\Omega_{\rm M}=0.286$ and  $\Omega_\Lambda=0.714$ \citep{2016A&A...594A..13P} and adopt the convention $Q_{\rm x}=Q/10^{\rm x}$ in c.g.s. units.

\section{Closure Relations in a stratified environment with variation of microphysical parameters}
\label{sec2}

Two types of shocks may be generated by the expansion of a relativistic jet into the circumburst medium: the forward shock, which travels outward, and the reverse shock, which travels inward, inside the jet. Taking into account the hydrodynamic of the forward-shock evolution with the Blandford-Mckee solution \citep{1976PhFl...19.1130B}, the bulk Lorentz factor becomes $ \Gamma = 137.9\left(\frac{1+z}{1.022}\right)^{\frac{3-k}{2(4-k)}} A_{\rm k,-1}^{-\frac{1}{2(4-k)}} E_{53}^{\frac{1}{2(4-k)}} t_2^{\frac{k-3}{2(4-k)}}$, where $E$ is the isotropic-equivalent kinetic energy, and $(1+z)$ is the correction due to the redshift.   We consider a profile of the circumburst density  $n(r) =A_{\rm k_s} r^{\rm -k}$, with $0\leq k < 3$ and the evolution of microphysical parameters to accelerate electrons and amplify the magnetic field as $\epsilon_{\rm e}= \varepsilon_{\rm e} \left(t/t_0\right)^{-\rm a}$ and $\epsilon_{\rm B}=\varepsilon_{\rm B} \left(t/t_0\right)^{-\rm b}$, respectively. The term $t_0$ corresponds to the onset of the variation of the microphysical parameters, which we will set to $10^2\,{\rm s}$ with $a=-0.5$ and $b=0.5$, and we will use ${\rm k=2}$ for $A_{\rm k_s}=A_{\rm k}\frac{\dot{M}_{\rm W}}{4\pi m_p v_{\rm W}}=A_{\rm k}3.0\times 10^{35}\,{\rm cm^{-1}}$,\footnote{The terms $v_{\rm W}$ and $\dot{M}_{\rm W}$ correspond to the wind velocity and the mass-loss rate, respectively, and $m_p$ is the proton mass.} unless otherwise indicated.  Accelerated electrons are distributed in accordance with their Lorentz factors ($\gamma_e$) and are described by the electron power index ($p$) as $N(\gamma_e)\,d\gamma_e \propto \gamma_e^{-p}\,d\gamma_e$ for $\gamma_m\leq \gamma_{\rm e}$, where $\gamma_m$ is the minimum electron Lorentz factor.  The comoving magnetic field in the blastwave $B'^2/(8\pi)=\varepsilon_B\,U_B$  can be obtained from the energy density $U_B=[(\hat\gamma\Gamma +1)/(\hat\gamma - 1)](\Gamma -1)n(r) m_pc^2$ where $\hat\gamma$ is the adiabatic index \citep{1999MNRAS.309..513H}, and $c$ is the speed of light.   Given the electron energy density, and the synchrotron, dynamical and the acceleration timescales, the minimum, cooling and the maximum electron Lorentz factors become   

{\small
\bary
\label{gam_m_w}
\gamma_{\rm m}&=& 2.9\times 10^3\, g_p\left(\frac{1+z}{1.022}\right)^{\frac{3-k}{2(4-k)}}\, \,\varepsilon_{e,-1} \,  A_{\rm k,-1}^{-\frac{1}{2(4-k)}}\, E_{53}^{\frac{1}{2(4-k)}}\, t_{2}^{-\frac{3+8a-k(1+2a)}{2(4-k)}},\cr
\gamma_{\rm c}&=&2.1\times 10^2\,(1+Y)^{-1} \left(\frac{1+z}{1.022}\right)^{-\frac{1+k}{2(4-k)}}\,\varepsilon^{-1}_{B,-2}    A_{\rm k,-1}^{-\frac{5}{2(4-k)}}E_{53}^{-\frac{3-2k}{2(4-k)}}\, t_{2}^{\frac{1+8b+k(1-2b)}{2(4-k)}},\cr
\gamma_{\rm max}&=& 1.5\times 10^7\,\left(\frac{1+z}{1.022}\right)^{-\frac{3}{4(4-k)}}\,  \varepsilon^{-\frac14}_{B,-2}\,A_{\rm k,-1}^{-\frac{3}{4(4-k)}}\, E^{-\frac{1-k}{4(4-k)}}_{53}\,  t_{2}^{\frac{3+b(4-k)}{4(4-k)}}\,,\hspace{6.cm}
\eary
}
%
respectively, where $g_p=\frac{p-2}{p-1}$ and $Y$ is the Compton parameter \citep{2001ApJ...548..787S, 2010ApJ...712.1232W}.  Given the comoving magnetic field, the bulk and electron Lorentz factors, the synchrotron spectral breaks  $\nu^{\rm syn}_{\rm i}=\frac{hq_e}{2\pi m_ec}(1+z)^{-1}\Gamma\gamma^{2}_{\rm i}B'$ with ${\rm i=m, c}$ become

{\small
	\bary
 \label{nums}
	h\nu_{\mathrm{m}}^{\rm syn} &=&   10.9\,{\rm eV}\,  g_p^{2} \left(\frac{1+z}{1.022}\right)^{\frac{1}{2}}\varepsilon_{\rm e,-1}^{2}\varepsilon_{\rm B,-2}^{\frac{1}{2}}E_{53}^{\frac{1}{2}}t_2^{-\frac{3+4a+b}{2}}, \cr
h\nu_{\mathrm{c}}^{\rm syn} &=&  5.5\times 10^{-2}\,{\rm eV}\, \left(\frac{1+z}{1.022}\right)^{-\frac{k+4}{2(4-k)}}A_{\rm k,-1}^{-\frac{4}{4-k}}(1+Y)^{-2}\varepsilon_{\rm B,-2}^{-\frac{3}{2}}E_{53}^{\frac{3k-4}{2(4-k)}}t_2^{\frac{3k+3b(4-k)-4}{2(4-k)}}. \hspace{4.9cm}
\eary
}

The terms $h$, $m_e$ and $q_e$ correspond to the Planck constant, electron mass and the elementary charge, respectively.  Using the synchrotron radiation power per electron $P_{\nu_i}\simeq \frac{\sqrt{3}q_e^3}{m_ec^2}(1+z)^{-1}\Gamma B'$ and the total number of emitting electrons $N_e=\frac{4\pi}{3-k} n(r)  r^3$ with $r\simeq \frac{2\Gamma^2 c}{(1+z)} t$ the shock radius, the maximum flux of synchrotron radiation $F^{\rm syn}_{\rm max}=\frac{(1+z)^2}{4\pi D_z^2}N_eP_{\nu_i}$, becomes

		\begin{eqnarray}\label{syn_flux}
			F_{\mathrm{\rm max}}^{\rm syn} &=&  86.2\,{\rm mJy}\,\left(\frac{1+z}{1.022}\right)^{\frac{16-3k}{2(4-k)}}A_{\rm k,-1}^{\frac{2}{4-k}}D_{\rm z,26.5}^{-2}\varepsilon_{\rm B,-2}^{\frac{1}{2}}E_{53}^{\frac{8-3k}{2(4-k)}}t_2^{\frac{b(k-4)-k}{2(4-k)}}\,, \hspace{6.2cm}
	\end{eqnarray}

where {\small $d_{\rm z}=(1+z)\frac{c}{H_0}\int^z_0\,\frac{d\tilde{z}}{\sqrt{\Omega_{\rm M}(1+\tilde{z})^3+\Omega_\Lambda}}$}  \citep{1972gcpa.book.....W}  is the luminosity distance.   Taking into account the evolution of spectral breaks and maximum flux of the synchrotron forward-shock model with variation of the microphysical parameters (Eqs. \ref{nums} and  \ref{syn_flux}), we derive the CRs shown in Table \ref{Table1}.    For the case of $a=0$ and $b=0$, the microphysical parameters do not evolve, and the synchrotron forward-shock model and CRs derived in \cite{1998ApJ...497L..17S, 2000ApJ...536..195C, 1998ApJ...503..314P} are recovered for the constant-density ($k=0$) and stellar-wind ($k=2$) environments.\\

When the same set of electrons that emits synchrotron radiation also up-scatters those photons to higher energy, we have the SSC mechanism. Given the electron Lorentz factors (Eq. \ref{gam_m_w}) and synchrotron spectral breaks (Eq. \ref{nums}), the SSC spectral breaks $h\nu^{\rm ssc}_{\rm i}\simeq \gamma^2_{\rm i} h\nu^{\rm syn}_{\rm i}$ can be written as \citep[e.g., see][]{2001ApJ...548..787S}

{\small

	\bary
    \label{nums_ssc}
			h\nu_{\mathrm{m}}^{\rm ssc} &=& 1.9\times 10^{-1}\,{\rm GeV}\,g_p^{4} \left(\frac{1+z}{1.022}\right)^{\frac{10-3k}{2(4-k)}}A_{\rm k,-1}^{-\frac{1}{4-k}}\varepsilon_{\rm e,-1}^{4}\varepsilon_{\rm B,-2}^{\frac{1}{2}}E_{53}^{\frac{6-k}{2(4-k)}}t_2^{\frac{bk+8ak+5k-4b-32a-18}{2(4-k)}}, \cr
  	h\nu_{\mathrm{c}}^{\rm ssc} &=& 4.8\times 10^{-6}\,{\rm GeV}\, \left(\frac{1+z}{1.022}\right)^{-\frac{3(k+2)}{2(4-k)}}A_{\rm k,-1}^{-\frac{9}{4-k}}(1+Y)^{-4}\varepsilon_{\rm B,-2}^{-\frac{7}{2}}E_{53}^{\frac{7k-10}{2(4-k)}}t_2^{\frac{5k+7b(4-k)-2}{2(4-k)}}. \hspace{6.5cm}
	\eary
}

Given the maximum synchrotron flux (Eq. \ref{syn_flux}),  the maximum flux of the SSC process  $F^{\rm ssc}_{\rm max}\sim \sigma_T n(r)\,r\, F^{\rm syn}_{\rm max}$ yields
	
	\begin{eqnarray}\label{ssc_flux}
			F_{\mathrm{\rm max}}^{\rm ssc} &=& 3.6\times 10^{-4}\,{\rm mJy}\, g_p^{-1}\,\left(\frac{1+z}{1.022}\right)^{\frac{14-k}{2(4-k)}}A_{\rm k,-1}^{\frac{5}{4-k}}D_{\rm z,26.5}^{-2}\varepsilon_{\rm B,-2}^{\frac{1}{2}} E_{53}^{\frac{5(2-k)}{2(4-k)}}t_2^{\frac{2-b(4-k)-3k}{2(4-k)}}\,.\hspace{6.2cm}
	\end{eqnarray}	

Taking into account the evolution of the spectral breaks and maximum flux of the SSC forward-shock model with variation of the microphysical parameters (Eqs. \ref{nums_ssc} and  \ref{ssc_flux}), we derive the CRs shown in Table \ref{Table2}.    For the case of $a=0$ and $b=0$, the microphysical parameters are constant, and the SSC forward-shock model including CRs presented in \cite{2022ApJ...934..188F} is recovered for the constant-density ($k=0$) and stellar-wind ($k=2$) environments.\\

\subsection{Hard spectral index ($1\leq p \leq 2$)}

For an electron distribution with a hard spectral index ($1\leq p \leq 2$), the minimum Lorentz factor is  $\gamma_{\rm m}= \left[\frac{m_p}{m_e}\tilde{g}_p \epsilon_e \Gamma \gamma_{\rm max}^{p-2} \right]^{\frac{1}{p-1}}$  with $\tilde{g}_p=\frac{2-p}{p-1}$ \citep[e.g., see][]{2001ApJ...558L.109D}.  Given the bulk Lorentz factor and the maximum energy that electrons can radiate, the minimum Lorentz factor becomes

{\small
\be
\label{gam_m_w_p}
\gamma_{\rm m}=
7.4\times 10^2\,\tilde{g}_p^{\frac{1}{p-1}}\,\left(\frac{1+z}{1.022}\right)^{\frac{3(4-p)-2k}{4(4-k)(p-1)}}\,  \varepsilon^{\frac{1}{p-1}}_{e,-1}  \varepsilon^{\frac{2-p}{4(p-1)}}_{B,-2}  A_{\rm k,-1}^{\frac{4-3p}{4(4-k)(p-1)}}E_{53}^{\frac{4-p-k(2-p)}{4(4-k)(p-1)}}\, t_2^{-\frac{12+4a(4-k)-2k+b(4-k)(2-p)-3p}{4(4-k)(p-1)}}\,.
\ee
}

Eq. \ref{gam_m_w_p} is different from the minimum electron Lorentz factor (Eq. \ref{gam_m_w}) usually derived for $p>2$. Consequently,  the characteristic spectral break  $\nu^{\rm syn}_{\rm m}=\frac{hq_e}{2\pi m_ec}(1+z)^{-1}\Gamma\gamma^{2}_{\rm i}B'$ for the synchrotron forward-shock model becomes

{\small
	\be
 \label{nums_p}
	h\nu_{\mathrm{m}}^{\rm syn}= 0.67\,{\rm eV}\,\tilde{g}_p^{\frac{2}{p-1}}\left(\frac{1+z}{1.022}\right)^{\frac{14+k(p-3)-5p}{2(4-k)(p-1)}}A_{\rm k,-1}^{\frac{2-p}{2(4-k)(p-1)}}\varepsilon_{\rm e,-1}^{\frac{2}{p-1}}\varepsilon_{\rm B,-2}^{\frac{1}{2(p-1)}} E_{53}^{\frac{p+2-k}{2(4-k)(p-1)}}t_2^{\frac{(4a+b)(k-4)+k-3p+kp-6}{2(4-k)(p-1)}}\,.
\ee
}

For $a=0$ and $b=0$, the typical characteristic spectral breaks evolving as $\nu_{\mathrm{m}}^{\rm syn}\propto t^{-\frac{3(p+2)}{8(p-1)}}$ and $\propto t^{-\frac{p+4}{4(p-1)}}$ for constant and wind medium, respectively, are recovered \citep{2001ApJ...558L.109D}. In this case, the SSC spectral break $h\nu^{\rm ssc}_{\rm i}\simeq \gamma^2_{\rm i} h\nu^{\rm syn}_{\rm i}$ evolving during the afterglow becomes

{\small
	\bary
    \label{nums_ssc_p}
			h\nu_{\mathrm{m}}^{\rm ssc}= 
			 7.4\times 10^{-4}\,{\rm GeV}\,\tilde{g}_p^{\frac{4}{p-1}} \left(\frac{1+z}{1.022}\right)^{\frac{26+k(p-5)-8p}{2(4-k)(p-1)}}A_{\rm k,-1}^{\frac{3-2p}{(4-k)(p-1)}}\varepsilon_{\rm e,-1}^{\frac{4}{p-1}}\varepsilon_{\rm B,-2}^{\frac{3-p}{2(p-1)}}E_{53}^{\frac{6+k(p-3)}{2(4-k)(p-1)}}t_2^{\frac{8a(k-4)-18+3k-b(k-4)(p-3)+kp}{2(4-k)(p-1)}}.
	\eary
}

Tables \ref{Table1} and \ref{Table2} show the CRs of synchrotron and SSC forward-shock model, respectively,  for a hard spectral index ($1\leq p \leq 2$).   For the case of $a=0$ and $b=0$, the microphysical parameters are constant, and the synchrotron CRs presented in \cite{2001ApJ...558L.109D} are recovered for the constant-density ($k=0$) and stellar-wind ($k=2$) environments.

\section{\textit{Fermi}-LAT GRB Catalog and Closure Relations}
\label{sec3}

\subsection{The methodical approach}

We chose a sample of 86 LAT-detected bursts from the extensive list provided in 2FLGC \citep{2019ApJ...878...52A}.  The duration of the high-energy emission from these bursts, which was temporally extended, was between 31 seconds (for GRB 141102) and 34,366 seconds (for GRB 130427).   In 2FLGC, the temporal emission is described with a SPL function; $F(t)\propto t^{-\alpha_{\rm LAT}}$, or a BPL: $F(t)\propto t^{-\alpha_{\rm LAT,1}}$ for $t<t_{\rm br, LAT}$ and $\propto t^{-\alpha_{LAT,2}}$ for $t_{\rm br, LAT}< t$ with $t_{\rm br, LAT}$ the break time. Similarly, the spectral emission is reported through the photon index $\Gamma_{\rm LAT}=\beta_{\rm LAT}+1$. For this analysis, we follow our previous strategy exhibited in \cite{2021PASJ...73..970D, 2021ApJS..255...13D, 2020ApJ...903...18S}.  The uncertainties displayed in Figures \ref{Fig3} and \ref{Fig4} are in the 1$\sigma$ range, and we take into account the uncertainties among the $\alpha$ and $\beta$ values to be dependent; thus, these figures are displayed  with ellipses. We require the temporal and spectral indices determined by the PL function $F_{\rm \nu} \propto t^{-\alpha} \nu^{-\beta}$ to examine the CRs. We use the temporal and spectral indexes from the 2FLGC.  Using the CRs of the synchrotron and SSC afterglow model derived and listed in Tables \ref{Table1} and  \ref{Table2},  we investigate if these CRs satisfy the spectral and temporal index  of those bursts reported in 2FLGC.

\begin{landscape}
\begin{figure*}
    \centering
    \includegraphics[width=1.2\textwidth]{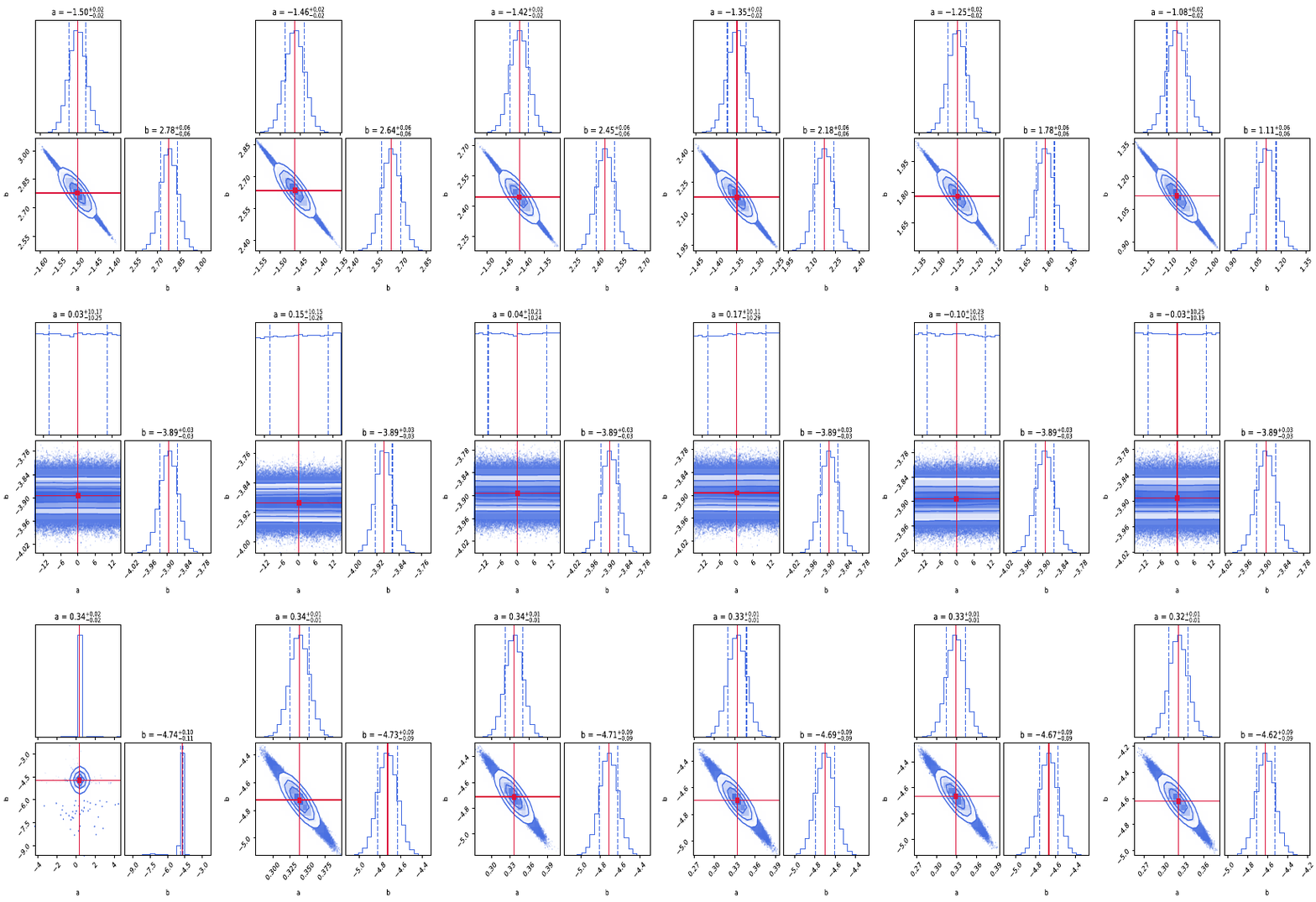}
    \caption{Corner plots of MCMC parameter estimation ($a$ and $b$) are shown for synchrotron CRs for different circumbust environments. The first column corresponds to a homogeneous medium with ${\rm k=0}$, the second assumes $\rm k=0.5$ and so on, with the last column taking the value ${\rm k=2.5}$. The rows, from top to bottom, are in accordance with the cooling condition  ${\rm \nu_m^{sync} < \nu_{\rm LAT} < \nu_c^{sync} }$,  ${\rm \nu_c^{sync} < \nu_{\rm LAT} < \nu_m^{sync} }$ and ${\rm max\{\nu_m^{sync},\nu_c^{sync} \} < \nu_{\rm LAT}}$, respectively. The histograms show the marginalized posterior densities with the median values in red lines.}
    \label{Fig1}
\end{figure*}
\end{landscape}

Figure \ref{Fig1} shows the corner plots of Markov Chain Monte Carlo (MCMC) parameter estimation ($a$ and $b$) in the scenario of the synchrotron CRs. Table \ref{Table3} displays the results from MCMC simulations of the synchrotron afterglow model evolving in a stratified circumburst medium. Each column in Figure \ref{Fig1} corresponds to a different circumburst environment, with the first column taking the value ${\rm k=0}$ and each subsequent column having a $0.5$ increment from the previous one. The rows follow a different cooling condition. The first row is in the slow cooling regime (${\rm \nu_m^{syn} < \nu_{\rm LAT} < \nu_c^{syn} }$), the second is in the fast cooling regime (${\rm \nu_c^{sync} < \nu_{\rm LAT} < \nu_m^{sync} }$) and the third one is for the condition ${\rm max\{\nu_m^{syn},\nu_c^{syn} \} < \nu_{\rm LAT}}$. Each histogram shows the marginalized posterior densities with the median values in red lines. Focusing on the first row, we notice that, in general, $a$ takes negative values and $b$ positive ones. This means that for our sample of GRBs, the electron microphysical parameter increases in time while the magnetic one decreases. By comparing each column in this row, we note that as the value of $\rm k$ is increased, the magnitude of the power indexes decreases; namely, for $\rm k=0$, we have $(a,b)=(-1.50,2.77)$ and for $\rm k=2.5$ we find $(a,b)=(-1.08,1.11)$, so as the stratification parameter increases, the time evolution of the microphysical parameters becomes weaker. Regarding the third row, we find that the behavior is the opposite, namely that $a$ takes positive values and $b$ negative ones. We also notice that variation of the stratification in this cooling condition leads to a negligible variation of the best-fit values of the power indexes, especially for $a$, as we find that for $\rm k=0$, we have $(a,b)=(0.34,-4.74)$ and for $\rm k=2.5$ we find $(a,b)=(0.32,-4.62)$.   Finally, the parameter estimation in the second row, which corresponds to the fast cooling regime, is not valid, as the corner plots show no clear region where the index $a$ is extreme. In this case, no further analysis is possible.

\begin{landscape}
\begin{figure*}
    \centering
    \includegraphics[width=1.2\textwidth]{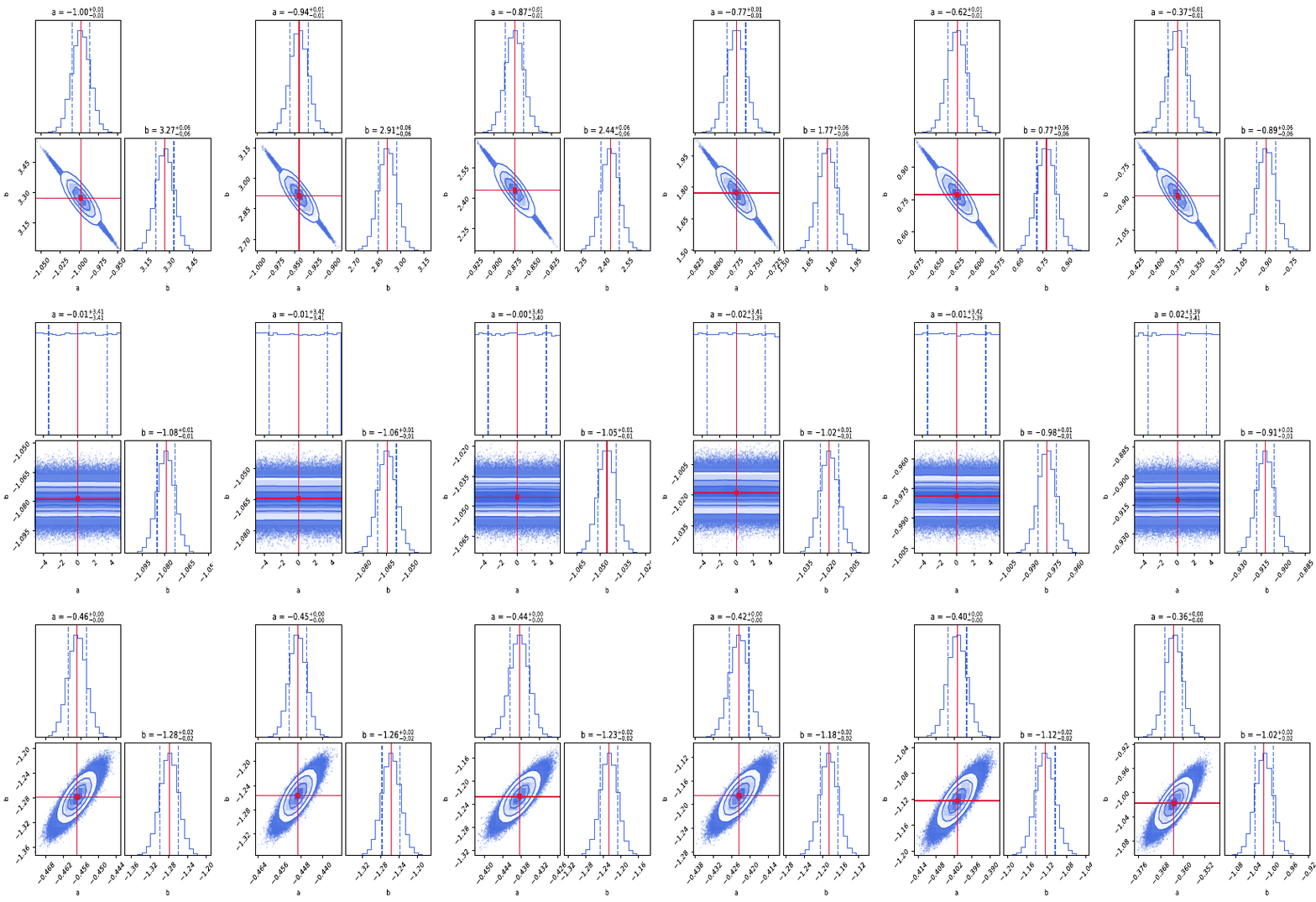}
    \caption{The same as Figure \ref{Fig1}, but for SSC process.}
    \label{Fig2}
\end{figure*}
\end{landscape}

\begin{figure*}
    \centering
    \includegraphics[width=1.\textwidth]{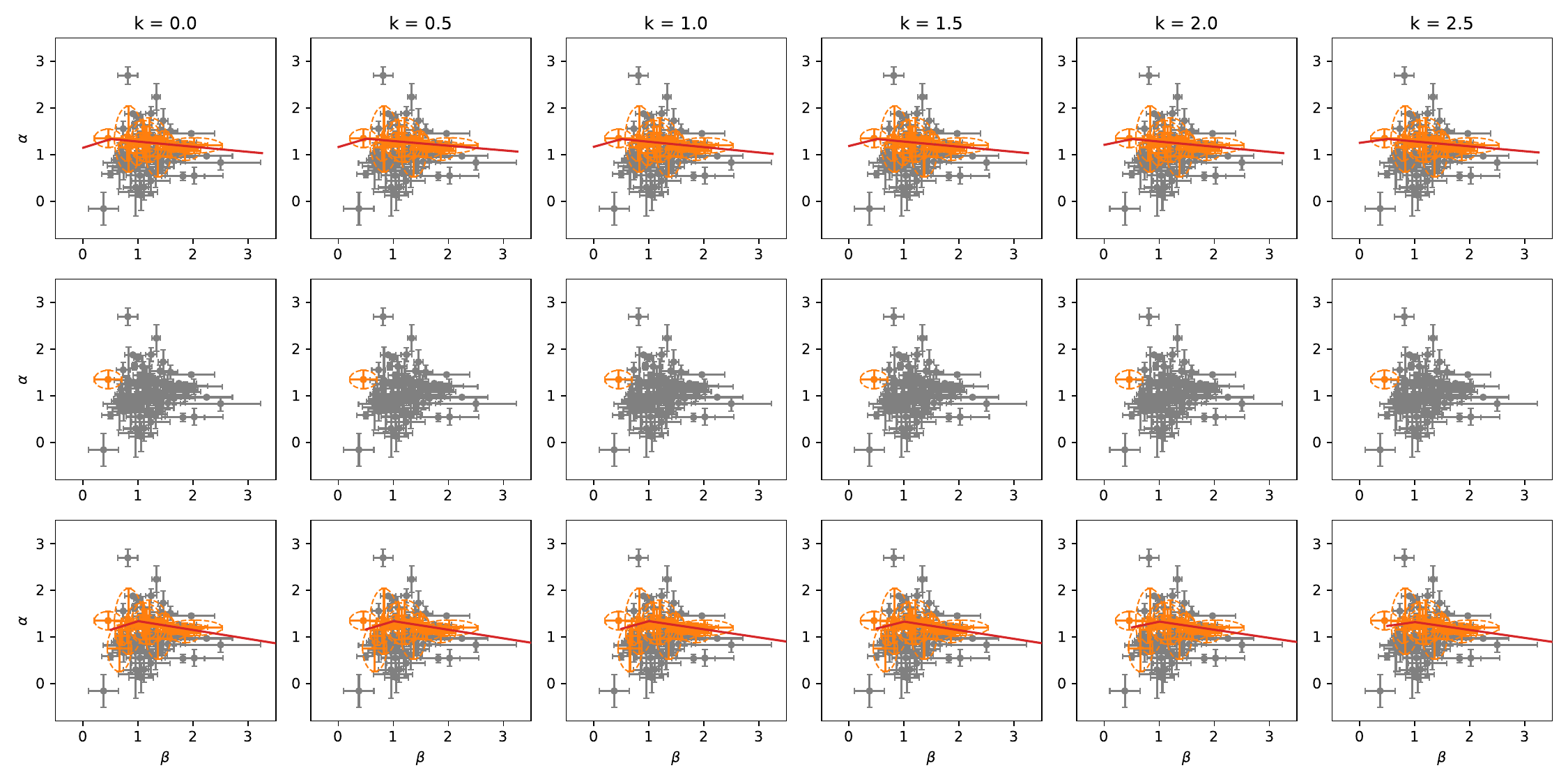}
    \caption{Synchrotron CRs  for circumbust environments with ${\rm k=0}$, $0.5$, $1.0$, $1.5$, $2.0$ and $2.5$. From top to bottom corresponds to the cooling condition  ${\rm \nu_m^{sync} < \nu_{\rm LAT} < \nu_c^{sync} }$,    ${\rm \nu_c^{sync} < \nu_{\rm LAT} < \nu_m^{sync} }$  and ${\rm max\{\nu_m^{sync},\nu_c^{sync} \} < \nu_{\rm LAT}}$, respectively. The orange ellipses display the cases in which the CR are satisfied within 1 sigma error bars for the best-fit values reported in Table \ref{Table3}. The red lines and dots correspond to the relationships themselves.}
    \label{Fig3}
\end{figure*}

\begin{figure*}
    \centering
    \includegraphics[width=1.\textwidth]{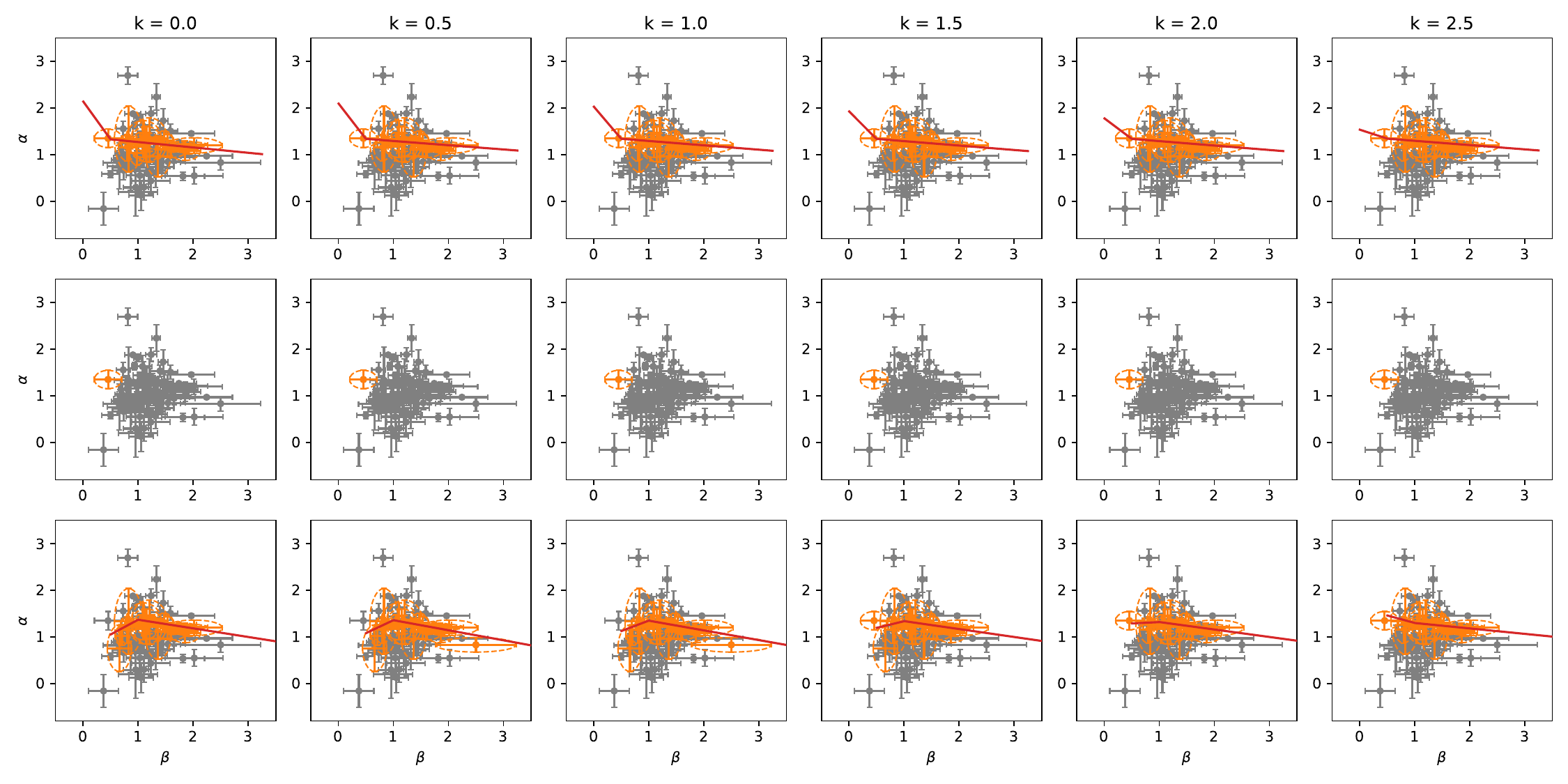}
    \caption{The same as Figure \ref{Fig3}, but for SSC process.}
    \label{Fig4}
\end{figure*}

Figure \ref{Fig2} displays the corner plots of MCMC parameter estimation ($a$ and $b$) in the case of the SSC CRs.  Table \ref{Table3} displays the results from MCMC simulations of the SSC afterglow model evolving in a stratified circumburst medium. Figure \ref{Fig2} follows the same conventions as the ones in Figure \ref{Fig1}, with the sole difference being that the cooling frequencies are the SSC frequencies instead of the synchrotron ones.   In the case of the SSC CRs, we generally find that the best-fit parameters are substantially different from the ones in the synchrotron scenario. We also find that the second row of this Figure suffers the same problem as the second row of Figure \ref{Fig1}, so no parameter estimation is possible in this case.    In the case of the first row, $a$ takes negative values, but as $\rm k$ increases, they become more positive. On the other hand, $b$ takes positive values for $0\leq k\leq2.0$, and as $\rm k$ increases, they decrease until a change of sign for $\rm k=2.5$. For the final row, we notice that similar to Figure \ref{Fig1}, variation in $\rm k$ does not substantially change the best-fit values of $a$ and $b$. However, we notice that in this cooling condition, both indexes are negative, which is a stark contrast to the synchrotron case in which the indexes had different signs.


Figure \ref{Fig3} shows the CRs of the synchrotron forward-shock model evolving in a stratified medium, from top to bottom $k=0$, $0.5$, $1$, $1.5$, $2$ and $2.5$, respectively, with the best-fit values obtained with our MCMC and listed in Table \ref{Table3}. The fast and slow regimes for each cooling condition are shown for $1<p<2$ and $p>2$. Table \ref{Table4} summarizes the number and percentage of bursts following each cooling condition for $1<p<2$ and $p>2$ and the profile index $k=0$, $0.5$, $1$, $1.5$, $2$ and $2.5$.  For the cooling condition ${\rm \nu_m^{syn} < \nu_{\rm LAT} < \nu_c^{syn} }$ the best case is most consistent with ${\rm k=0, 1.0, 1.5, 2}$ (23 GRBs, 27.06\%). For the cooling condition ${\rm \nu_c^{syn} < \nu_{\rm LAT} < \nu_m^{syn} }$, all cases of ${\rm k}$ are equally likely (1 GRB, 1.18\%).  For the cooling condition ${\rm max\{\nu_m^{syn},\nu_c^{syn} \} < \nu_{\rm LAT}}$, the best case is most consistent with ${\rm k=1}$ (24 GRBs, 28.24\%) and worst scenario is for ${\rm k=2.5}$ (21 GRBs, 24.71\%).

Figure \ref{Fig4} shows the CRs of the SSC forward-shock model evolving in a stratified medium, from top to bottom $k=0$, $0.5$, $1$, $1.5$, $2$ and $2.5$, respectively, with the best-fit values obtained with our MCMC and listed in Table \ref{Table3}. The fast and slow regimes for each cooling condition are shown for $1<p<2$ and $p>2$. Table \ref{Table5} summarizes the number and percentage of bursts following each cooling condition for $1<p<2$ and $p>2$ and  the profile index $k=0$, $0.5$, $1$, $1.5$, $2$ and $2.5$. For the cooling condition ${\rm \nu_m^{ssc} < \nu_{\rm LAT} < \nu_c^{ssc} }$ the best case is most consistent with ${\rm k=0}$ (23 GRBs, 27.06\%) followed by all cases with 22 GRBs (25.88\%).  For the cooling condition ${\rm \nu_c^{ssc} < \nu_{\rm LAT} < \nu_m^{ssc}}$, all cases of ${\rm k}$ are equally likely (1 GRB, 1.18\%).  For the cooling condition ${\rm max\{\nu_m^{ssc},\nu_c^{ssc} \} < \nu_{\rm LAT}}$, the best cases are most consistent with ${\rm k=0}$ and ${\rm k=0.5}$ (24 GRBs, 28.24\%).

We have obtained the synchrotron and SSC closure relations in the slow- and fast-cooling regimes, which provide the function $\alpha(\beta;a;b)$. We have noticed that not all cooling conditions are a function of $a$ and $b$ simultaneously, which lead to an inability to obtain a relation between both parameters in such cases. This, in turn, lead to a degeneracy in either $a$ or $b$, which appeared in our analysis as a huge uncertainty in our MCMC simulations.


\subsection{The light curves and Spectral Energy Distributions: Conditions for Plateaus and Peaks in 2FLGC}

Tables \ref{Table:sync} and \ref{Table:ssc} show the evolution of the synchrotron and SSC light curves with microphysical parameter variations, respectively, from an outflow that decelerates in a stellar wind ($k=2$) and constant-density ($k=0$) medium, and an electron distribution with spectral index in the range of $1<p<2$ and $p>2$. The evolution of temporal indexes is estimated using the synchrotron and SSC spectra for fast- and slow-cooling regimes together with the respective spectral breaks (Eqs. \ref{nums_ssc} and \ref{nums_ssc_p}) and the maximum flux (Eq. \ref{ssc_flux}).
 
\begin{figure*}
    \centering
    \includegraphics[width=1.\textwidth]{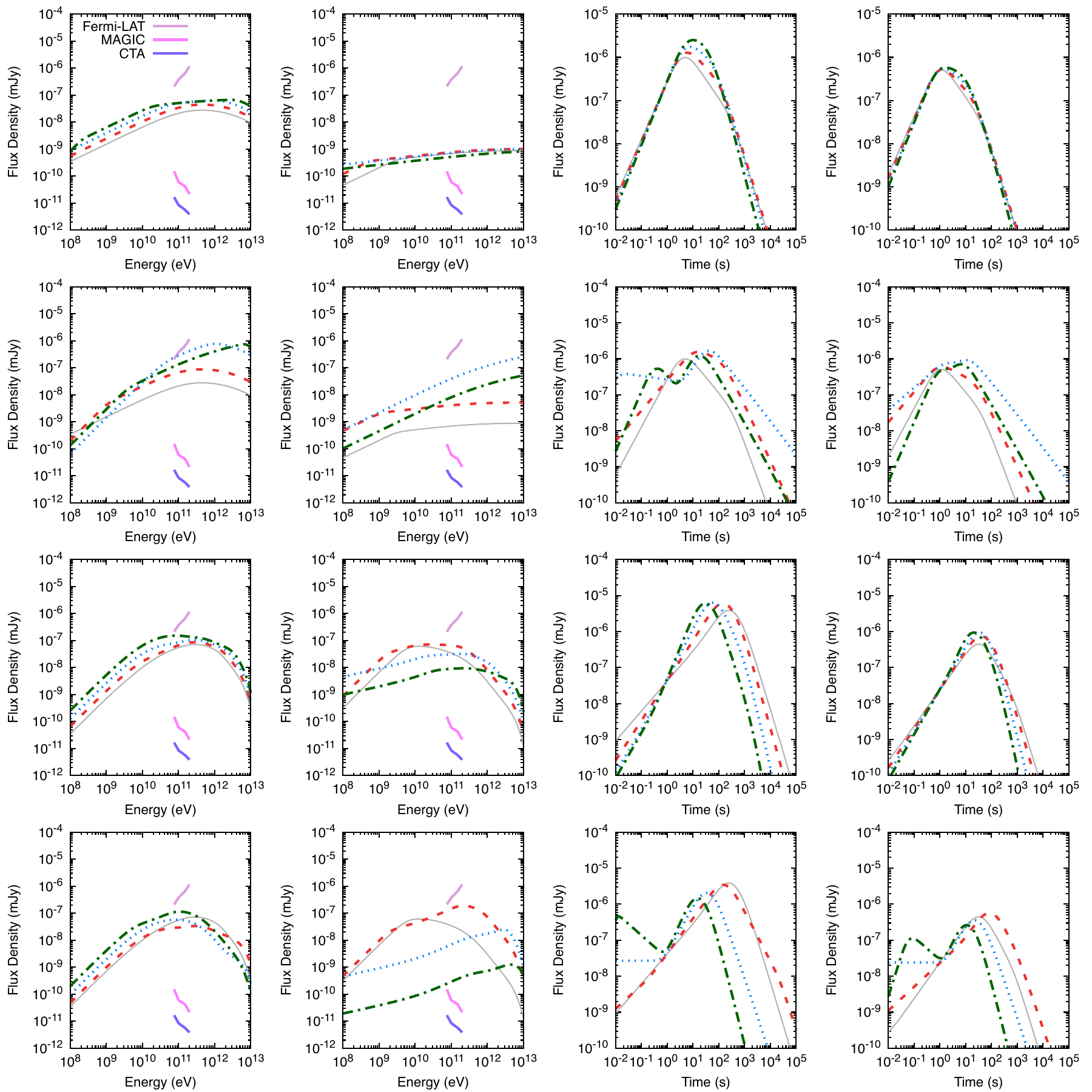}
    \caption{The SSC light curves (right) and spectra (left) in uniform-density afterglow model with variation of microphysical parameters. The SEDs are displayed at 10 (first column) and 100 s (second column), and the light curves for 1 (third column) and 10 (four column) GeV.   The SSC spectra are shown in the left-hand panels together with the CTA (Southern array, green line), MAGIC (purple line) and \textit{Fermi}-LAT (red line) sensitivities between 75 and 250 GeV at $3\times 10^4\,{\rm s}$ for a zenith angle of 20$^\circ$ \citep{2019ICRC...36..673F}.  The values of the parameters used are: $E=1\times 10^{52}\,{\rm erg}$, $n=10^{-1}\,{\rm cm^{-3}}$, $\varepsilon_{\rm e}=0.1$, $\varepsilon_{\rm e}=0.01$, $p(<2)=1.7$,  $p(2<)=2.4$ and $z=0.1$}
    \label{Fig6}
\end{figure*}

Figure \ref{Fig6} presents the SSC light curves and spectra obtained from the uniform-density afterglow model with a variation of microphysical parameters. The first and second columns show the spectra at 10 and 100 s, respectively, with the CTA, MAGIC, and \textit{Fermi}-LAT sensitivities in colored solid lines. The third and fourth columns display the light curves for 1 and 10 GeV, respectively. The first two rows have a value of the electron power index of $p=1.7$, while the last two have $p=2.4$. The first and third rows consider a variation of the $a$ parameter while keeping $b=0$ fixed. In this case, the colored lines correspond to $a=0$ (gray), $a=0.25$ (red), $a=0.5$ (blue) and $a=0.75$ (green).
On the other hand, the second and fourth rows keep $a=0$ fixed and vary the value of $b$. For these rows, the colored lines represent $b=0.25$ (gray), $b=0.5$ (red), $b=0.75$ (blue) and $b=1.25$ (green). Regarding the different spectra, we notice that variations in either $a$ or $b$ would be observable by MAGIC and CTA, but not by \textit{Fermi}-LAT as its sensitivity is not enough to capture a sufficient part of the spectra. In all panels, we can see that an increase in any of the microphysical parameters will, most of the time, lead to an increase in the flux density for small energy. However, this behavior depends on the energy, as we note in the panels in the last row, where the gray and red curves dominate for considerable energies. We also observe that differences in the power index of the microphysical parameters are not very apparent if the flux density is almost constant, as displayed in the panel in the first-row second column. In the case of the light curves, we note that variation in $a$ and $b$ will lead to changes in the cooling regimes, as exemplified by the panels in the second and last rows. Upon comparing the first and third rows (second and fourth), we also notice that variations in the $b$ parameter lead to more considerable changes in the light curves than variations in $a$.

\begin{figure*}
    \centering
    \includegraphics[width=1.\textwidth]{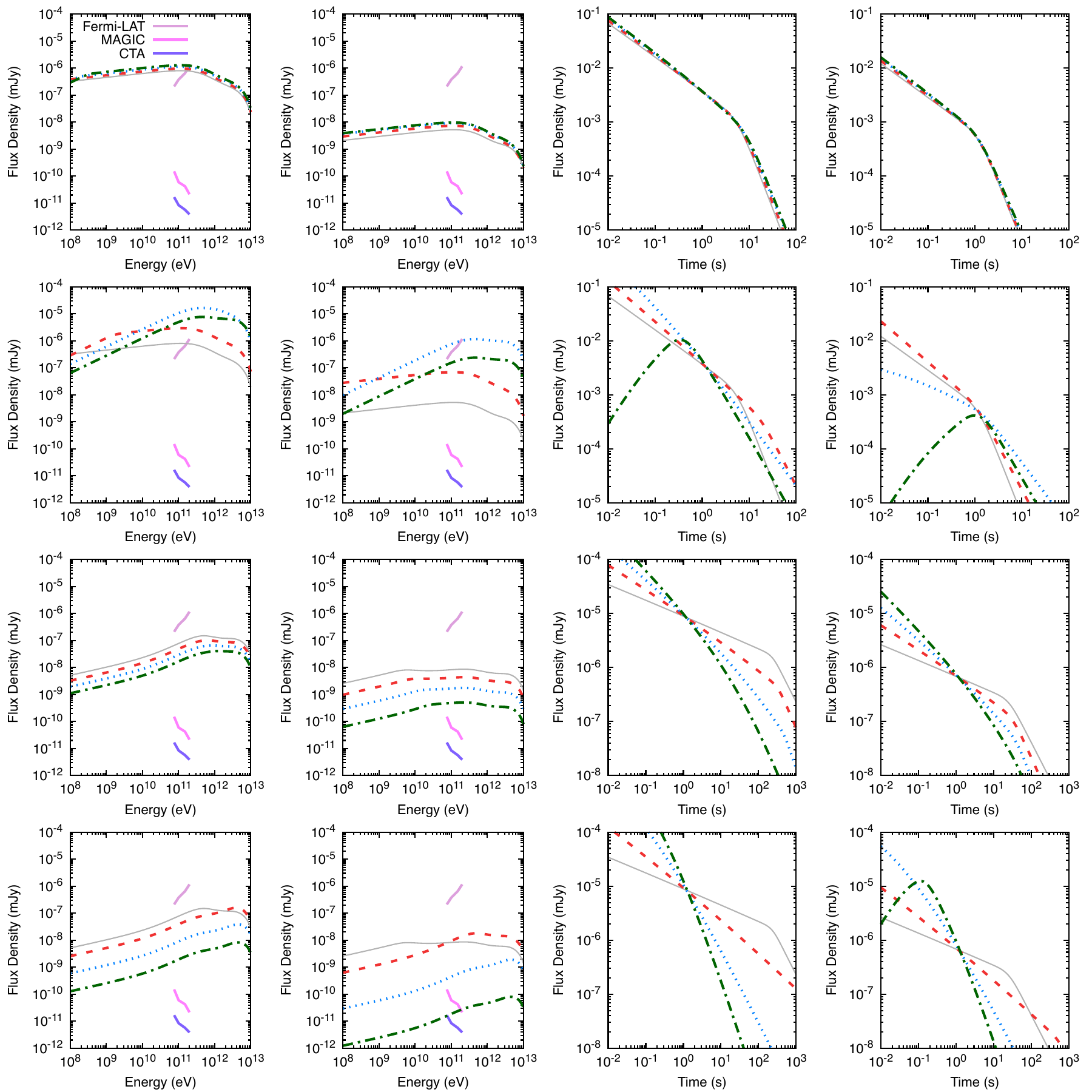}
    \caption{The same as Figure \ref{Fig6}, but for stellar-wind afterglow model with $A_{\rm W}=0.1$.}
    \label{Fig7}
\end{figure*}

Figure \ref{Fig7} is analogous to Figure \ref{Fig6}, with the only difference being that this Figure displays the SSC light curves and spectra obtained with the stellar-wind afterglow model. We note that the spectra predicted by the stellar-wind model are generally flatter than those in the constant-density medium. We also notice that for $2<p$, an increase in the power index of the microphysical parameters leads to less flux density, in contrast to the $1<p<2$ case and the whole constant-density model. In the case of the light curves, in the same manner as in Figure \ref{Fig6}, changes in $a$ and $b$ may lead to changes in the cooling conditions, an effect that is most apparent when varying $b$. Compared to the constant-density case, the wind model predicts more significant fluxes at early times but with steeper decays at later times. Additionally, we show that a plateau phase or a MeV-GeV peak in the LAT energy range caused by synchrotron and SSC processes could be expected depending on the values of $a$ and $b$.



\subsubsection{Parameter conditions for describing high-energy plateaus}

Like the X-ray afterglows recorded by the \textit{Swift}-XRT instrument, many GRBs display temporally prolonged high-energy emissions as measured by the \textit{Fermi}-LAT calorimeter. There is a late-time flattening in certain LAT light curves that is suggestive of X-ray plateaus.   \cite{2021ApJS..255...13D} investigated the presence of late-time flattening in LAT observations reported in 2FLGC, offering evidence that plateaus exist in the light curves.   They argued that the identical phenomenological model required to interpret X-ray plateaus might be used to describe some of the most energetic bursts; GRB 090510, GRB 090902B, and GRB 160509A, which exhibited gamma-ray photons above 10 GeV. The synchrotron and SSC flux could mimic a late-time flattening in any cooling condition of the light curves for both a constant-density and stellar-wind environment. This condition is fulfilled when the light curve's time dependence vanishes when the temporal variable's power index becomes zero.  For the synchrotron forward-shock model with $p>2$, the value of the parameter $b$ as a function of $a$ in this case becomes $b=1$, $b=-\frac{(3+4a)(p-1)}{p+1}$ ($=\frac{1 - 3p - 4a(p-1)}{p+1}$)  for $k=0$ ($k=2$) and $b=\frac{2 - 3p - 4a(p-1)}{p-2}$, for the cooling conditions ${\rm \nu_c^{syn} < \nu_{\rm LAT} < \nu_m^{syn} }$, ${\rm \nu_m^{syn} < \nu_{\rm LAT} < \nu_c^{syn} }$ and ${\rm max\{\nu_m^{syn},\nu_c^{syn} \} < \nu_{\rm LAT}}$, respectively. It is worth noting that the conditions for ${\rm \nu_c^{sync} < \nu_{\rm LAT} < \nu_m^{syn} }$ and ${\rm max\{\nu_m^{syn},\nu_c^{syn} \} < \nu_{\rm LAT}}$ are independent of the density profile.  Similarly, for the SSC forward-shock model, $b=-\frac{1}{10}$ ($=0$), $b=\frac{11 - 9p - 16a(p-1)}{2(p+1)}$ ($=-\frac{4p + 8a(p-1)}{p+1}$) and $b=\frac{10-9p - 16a(p-1)}{2(6-p)}$ ($=-\frac{4(2a+1)(p-1)}{6-p}$) for $k=0$ ($k=2$), for the cooling conditions ${\rm \nu_c^{ssc} < \nu_{\rm LAT} < \nu_m^{ssc} }$, ${\rm \nu_m^{ssc} < \nu_{\rm LAT} < \nu_c^{ssc} }$ and ${\rm max\{\nu_m^{ssc},\nu_c^{ssc} \} < \nu_{\rm LAT}}$, respectively. Therefore, with the appropriate values, the late-time flattening could explain GRB 090510, GRB 090902B, and GRB 160509A.

The parameter conditions for $a$ and $b$ in the synchrotron model and  the best-fit values of the parameter $a$ and $b$ shown in Table \ref{Table3} indicate that the cooling conditions for synchrotron afterglow scenario ${\rm \nu_c^{\rm ssc} < \nu_{\rm LAT} < \nu_c^{\rm ssc}}$ and ${\rm max\{\nu_m^{\rm syn},\nu_c^{\rm syn} \} < \nu_{\rm LAT}}$ could exhibit plateaus for both a homogeneous medium and a stellar wind. In this case, the fraction of energy given to the electrons decreases and  to the magnetic field increases, so that these variation could be associated with shell collisions. This turbulence distorts the field lines proving additional magnetic reconnections. It leads to the release of the accumulated magnetic field energy on the runway \citep[an ICMART\footnote{the Internal-Collision-induced MAgnetic Reconnection and Turbulence.} event;][]{2011ApJ...726...90Z}. It is worth noting that the no coincidence of neutrinos with GRBs \citep{2012Natur.484..351A, 2016ApJ...824..115A, 2015ApJ...805L...5A} could be explained in some magnetic dissipation scenarios such as the ICMART model \citep{2013PhRvL.110l1101Z}.   Similarly, the parameter conditions for $a$ and $b$ in the SSC model with the best-fit values of the parameter for SSC afterglow model shown in Table \ref{Table3} suggest that the cooling conditions ${\rm \nu_c^{\rm ssc} < \nu_{\rm LAT} < \nu_c^{\rm ssc}}$ and ${\rm max\{\nu_m^{\rm syn},\nu_c^{\rm syn} \} < \nu_{\rm LAT}}$ could exhibit high-energy plateaus.  For these these cooling conditions and with the values of $a$ and $b$, the fraction of energy given to accelerate electrons and amplified magnetic field increases at the same time, so that these variation could be associated with continuous energy injection with different efficiencies \citep{2006A&A...458....7I, 2006MNRAS.370.1946G}.



\subsubsection{Parameter conditions to explain high-energy peaks}

The 2FLGC exhibited  a subset of GRBs with a short-lasting  peak at the end of prompt episode and the beginning of the long-lasting emission \citep[e.g., see GRB 080916C, GRB 090510A, GRB 090902B, GRB 090926A, GRB 110731A, GRB 130427A and GRB 160509A][]{2009Sci...323.1688A, 2009ApJ...706L.138A, 2010ApJ...716.1178A, 2011ApJ...729..114A, 2013ApJ...763...71A, 2014Sci...343...42A}.   For instance,   GRB 160509A displayed a bright bump peaking at $\sim$ 20 s, followed by the long-lasting emission.  The synchrotron and SSC flux could mimic a short-lasting bump when the forward shock evolves in a constant-density and stellar-wind environment.

The increasing flux $t^{-\alpha}$ with $\alpha>-2$.  For the synchrotron forward-shock model with $p>2$, $b<9$, $b>-\frac{5 + 3p + 4a(p-1)}{p+1}$ ($>-\frac{7 + 3p + 4a(p-1)}{p+1}$) for $k=0$ ($k=2$) and $b>-\frac{3(p +2) + 4a(p-1)}{p-2}$, for the cooling conditions ${\rm \nu_c^{\rm syn} < \nu_{\rm LAT} < \nu_m^{\rm syn} }$, ${\rm \nu_m^{\rm syn} < \nu_{\rm LAT} < \nu_c^{\rm syn} }$ and ${\rm max\{\nu_m^{\rm syn},\nu_c^{\rm syn} \} < \nu_{\rm LAT}}$, respectively. It is worth noting that the conditions for ${\rm \nu_c^{\rm syn} < \nu_{\rm LAT} < \nu_m^{\rm syn} }$ and ${\rm max\{\nu_m^{\rm syn},\nu_c^{\rm syn} \} < \nu_{\rm LAT}}$ are independent of the density profile.  Similarly, for SSC forward-shock model, $b<\frac{3}{2}$ ($<\frac{8}{5}$), $b>-\frac{5 + 9p + 16a(p-1)}{2(p+1)}$ ($>-\frac{4[(p+2) + 2a(p-1)]}{p+1}$) and $b>-\frac{3(3p+2) + 16a(p-1)}{2(p-6)}$ ($-\frac{4[(p+1)+2a(p-1)]}{p-6}$) for $k=0$ ($k=2$), for the cooling conditions ${\rm \nu_c^{ssc} < \nu_{\rm LAT} < \nu_m^{ssc} }$, ${\rm \nu_m^{ssc} < \nu_{\rm LAT} < \nu_c^{ssc} }$ and ${\rm max\{\nu_m^{ssc},\nu_c^{ssc} \} < \nu_{\rm LAT}}$, respectively. Therefore, the high-energy peaks exhibited in particular bursts could be described by the variation of the microphysical parameters in the synchrotron and SSC scenario.  The conditions of $a$ and $b$ to emulate a bright peak can be satisfied when the instantaneous energy injection is supplied for a small time interval; so that the microphysical parameters only evolve during a short period. It is important mentioning that a superposition of synchrotron and SSC process could describe the high-energy peaks.

\subsection{Variation of Magnetic Parameter in Decaying Microturbulence}

In the GRB afterglow model, particle acceleration in a decelerating relativistic, collisionless shock front is an essential component \citep{1997ApJ...476..232M}. The long-standing questions regarding where the magnetic fields that penetrate the blast waves of the GRB originate and the mechanisms by which energy is transported from particles (e.g., protons and electrons) and magnetic fields in relativistic shocks have not been fully answered.   Relativistic collisionless shock front formation in a weakly magnetized surrounding medium is explained by the self-generation of intense small-scale electromagnetic fields mediating the transition between the upstream unshocked and the downstream shocked media.   Accelerated particles play a crucial role in setting off the microinstabilities that eventually create the self-generated field.  The self-generated microturbulence controls these particles' dispersion, which in turn guides the complexly non-linear acceleration process.  High-performance particle-in-cell (PIC) simulations have so far confirmed the validity of this entire approach \citep[e.g.,][]{2008ApJ...682L...5S, 2011ApJ...726...75S, 2009ApJ...698.1523S, 2009ApJ...695L.189M}.

The magnetic field amplification may occur in the GRB's afterglow shocks and is linked to the generation of turbulence at a small scale. The analysis listed in Table \ref{Table3} shows that the magnetic field is amplified only in the  case of slow cooling regime (${\rm \nu_m^{\rm j} < \nu < \nu_{c}^{\rm j}}$) for either synchrotron or SSC model. In this case, the parameters $a$ and $b$ lies in the range of  $-1.50\leq a\leq -1.08$ and $1.11\leq b\leq 2.77$  for  synchrotron  and $-1.00\leq a\leq -0.37$ and $0.77\leq b\leq 3.27$  for SSC. This Table also shows that the magnetic field is not amplified for $k=2.5$ in the SSC scenario.  It may be due to differences between two collisionless populations of particles \citep{1999ApJ...526..697M}. Due to the generation of microturbulence, the Fermi acceleration mechanism must be strongly influenced by it. The results of \cite{2006ApJ...645L.129L} suggest that Fermi acceleration is efficient when the magnetic field is amplified in scales shorter than the typical Larmor radius; in this case, turbulence helps particles gain energy by completing cycles through the shock front. The formation of a microturbulent layer guarantees the scattering of accelerated particles. This is the case of the cooling condition ${\rm max\{\nu_m^{ssc},\nu_c^{ssc} \} < \nu_{\rm LAT}}$ for SSC scenario where the values of $a$ and $b$ lie in the range of $-0.46\leq a\leq -0.36$ and $-1.28\leq b\leq -1.02$, as shown in Table \ref{Table3}.  In contrast, in the absence of this layer,  particles are advected away with a transverse magnetic field, suppressing the acceleration process. Table \ref{Table3} shows that this criterion is not satisfied for any cooling condition. Therefore, to understand how particle energy increases via Fermi acceleration and how it decreases via synchrotron radiation, it is crucial to know how quickly and how far the microturbulent layer decays \citep{2013MNRAS.428..845L}. The author considered the variation of the magnetic parameter $\propto t^{-b}$  and discussed the effects of the parameter $b$ on the resulting spectrum for synchrotron and dominant inverse Compton, identifying two scenarios.  In the first scenario, turbulence gradually decays with $-1<b<0$ and $b>-4/(p+1)$ for synchrotron and dominant inverse Compton, respectively, and in the second scenario, turbulence rapidly decays with $b<-1$ for synchrotron model and  $-3<b<-4/(p+1)$ for dominant inverse Compton.

The best-fit values of the parameter $b$ shown in Table \ref{Table3} indicate that only the cooling conditions ${\rm \nu_m^{\rm syn} < \nu_{\rm LAT} < \nu_c^{\rm syn} }$ and ${\rm \nu_m^{\rm ssc} < \nu_{\rm LAT} < \nu_c^{\rm ssc} }$ lie in the scenario of rapidly decaying microturbulence for both synchrotron and SSC processes.

\subsection{Bursts with a hard spectral index ($1\leq p \leq 2$) in 2FLGC}

Table \ref{tab:sample} shows the selection of bursts from 2FLGC with spectral index $\beta<1$ and temporal index $\alpha>1.3$, which can hardly be modeled with the CRs of the standard synchrotron forward-shock model. \cite{2019ApJ...883..134T} showed that the CRs of the slow-cooling regime ${\rm \nu_m^{syn} < \nu_{\rm LAT} < \nu_c^{syn} }$ for $2<p$ could satisfy the evolution of temporal and spectral index of these bursts, but only with the condition of having an atypical small value of $\epsilon_B<10^{-7}$. This condition must be considered so that the cooling spectral break is larger than the LAT energy band ($\nu_{\rm LAT} < \nu_c^{syn}$). \cite{2010MNRAS.403..926G} modeled the high-energy emission of 11 GRBs (up until October 2009) identified by the \textit{Fermi}-LAT. Based on the LAT observations, the authors determined that the relativistic forward shock was in the utterly radiative regime for a hard spectral index $p\sim 2$, $\sim t^{-\frac{10}{7}}$ instead of $t^{-1}$, as anticipated by the adiabatic synchrotron forward-shock model, being a slow-cooling regime in the constant-density medium the most favorable afterglow model.   The scenario of microphysical parameter variation could explain the evolution of this GRB sample's temporal and spectral indexes. For instance,  we can see that the only cooling conditions for the synchrotron model that satisfy these bursts would be ${\rm max\{\nu_m^{syn},\nu_c^{syn} \} < \nu_{\rm LAT}}$ for $1\leq p \leq 2$.

\begin{landscape}
\begin{figure*}
    \centering
    \includegraphics[width=1.3\textwidth]{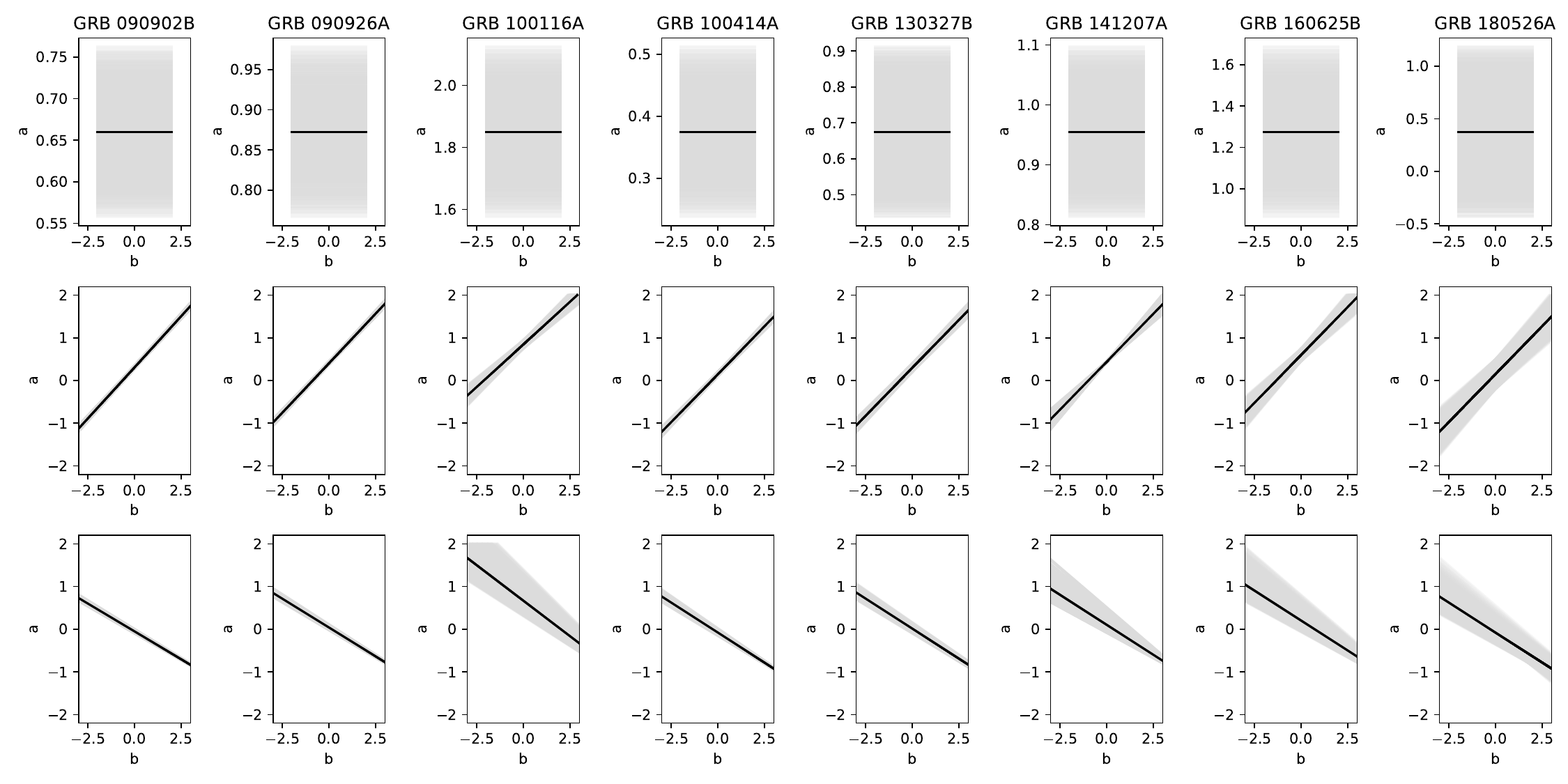}
    \caption{PL indices of microphysical variations for which the CRs are fulfilled for a constant-density medium.    The rows, from top to bottom, are in accordance with the cooling condition  ${\rm max\{\nu_m^{sync},\nu_c^{sync} \} < \nu_{\rm LAT}}$ for synchrotron model with $1<p<2$, ${\rm max\{\nu_m^{ssc},\nu_c^{ssc} \} < \nu_{\rm LAT}}$  and ${\rm \nu_m^{ssc} < \nu_{\rm LAT} < \nu_c^{ssc} }$ for SSC model with  $1<p<2$ and with $p>2$, respectively. }
    \label{Fig8}
\end{figure*}
\end{landscape}

For completeness, we now include the CRs of the SSC model in our analysis.   Figure \ref{Fig8} shows the PL indices of the microphysical parameters for which the CRs are fulfilled in the constant-density medium model for a sample of 8 GRBs. Each row corresponds to a different cooling condition. The first one is related to the cooling condition  ${\rm max\{\nu_m^{sync},\nu_c^{sync} \} < \nu_{\rm LAT}}$ for the synchrotron model with $1<p<2$, the second one to ${\rm max\{\nu_m^{ssc},\nu_c^{ssc} \} < \nu_{\rm LAT}}$ for the SSC model with  $1<p<2$ and the third one to ${\rm \nu_m^{ssc} < \nu_{\rm LAT} < \nu_c^{ssc} }$ for the SSC model with $p>2$. Each column corresponds to the particular GRB in question. In all panels, the solid line stands for the relation between the $a$ and $b$ parameters that satisfy the respective CR, while the gray region represents the propagated uncertainty from the uncertainty in the temporal and spectral PL indices from \citet{2019ApJ...878...52A}, summarized in Table \ref{tab:sample}.  In the first row, we bear in mind that the general behavior is maintained in all panels, namely that we can only find the best-fit $a$ parameter, but we cannot conclude on the $b$ parameter. This is easily explained by noticing that the appropriate CR in Table \ref{Table3} does not depend on $b$; as such, it remains unconstrained in our analysis. We also note that a positive $a$ is preferred for this cooling condition. Regarding the second row, we note that the best-fit value of $a$ is proportional to $b$, and the same relation seems to be valid for all GRBs except GRB 100116A, which prefers larger values of $a$. In the case of the third row, we see that the relation between $a$ and $b$ is now inversely proportional. In an analogous fashion to the second row, the same relation is approximately valid for all GRBs except GRB 100116A, which, as in the previous row, prefers larger values of $a$. It is essential to highlight that the case for $a=0$ and $b=0$ (standard synchrotron model) is obtained only in a few bursts when the uncertain regions are considered.

\begin{landscape}
\begin{figure*}
    \centering
    \includegraphics[width=1.3\textwidth]{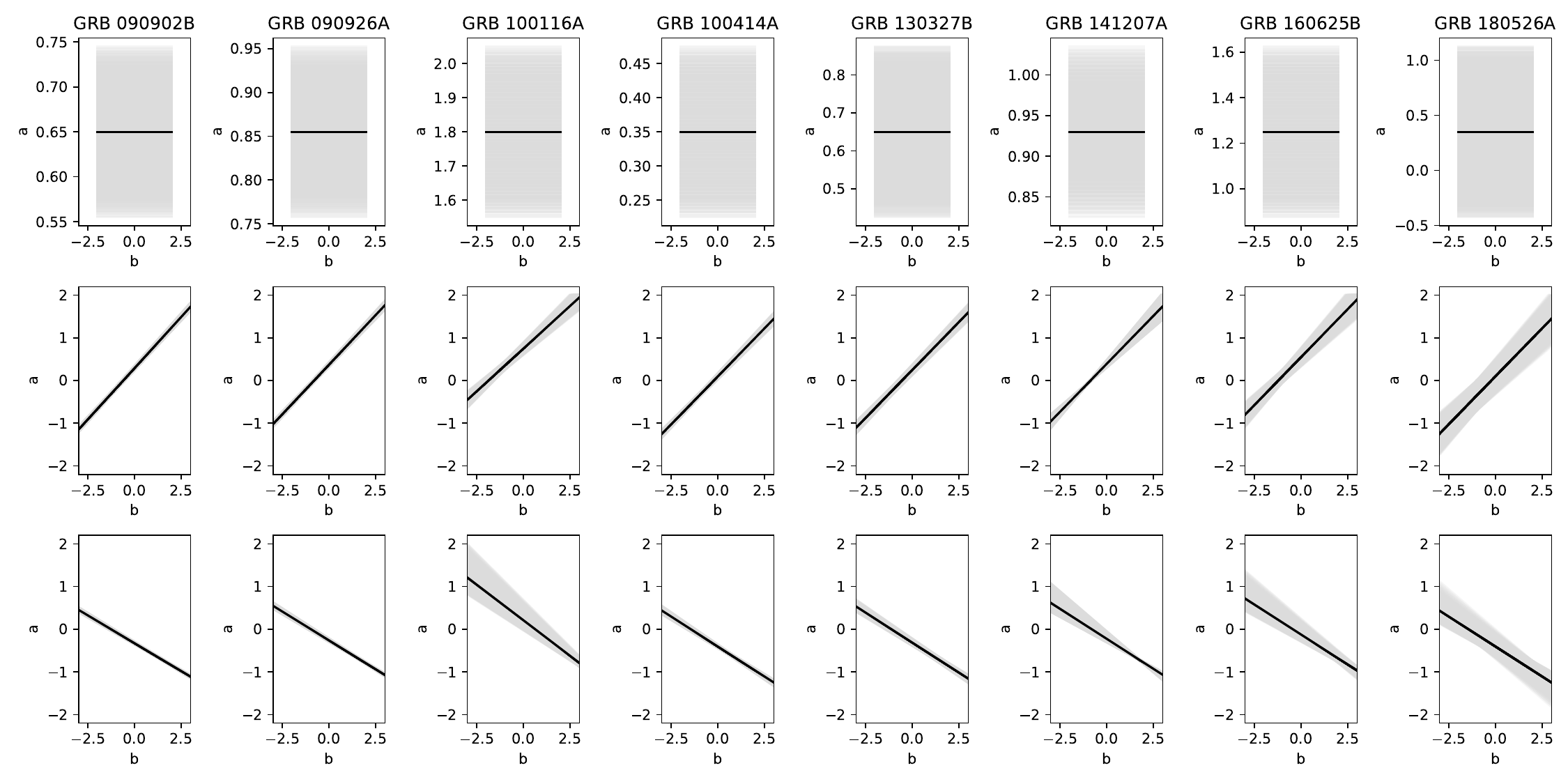}
    \caption{The same as Figure \ref{Fig8}, but for stellar-wind afterglow model.}
    \label{Fig9}
\end{figure*}
\end{landscape}

Figure \ref{Fig9} is analogous to Figure \ref{Fig8}, with the sole difference being that we consider the stellar-wind model here. The discussion for Figure \ref{Fig8} can be repeated for this one, as the same behavior is repeated in all panels. In all cases, all best-fit values are consistent between both Figures, so this type of analysis does not lead to a way to differentiate between a constant-density medium and a stellar wind.

\subsection{Stellar-wind/constant-density medium}

For both short and long-GRB progenitors, the external circumburst medium plays an essential role in the hydrodynamics of the ejected outflows.   A constant low-density medium is expected for sGRBs, which are originated from the merging of binary compact objects, such as a BH and a NS or NS-NS \citep{1992ApJ...392L...9D, 1992Natur.357..472U, 1994MNRAS.270..480T, 2011MNRAS.413.2031M}.  A stratified environment with different density profiles is identified for long GRBs, which are associated with the core collapse of massive stars \citep{1993ApJ...405..273W,1998ApJ...494L..45P}. 
In particular, a transition between a stratified and constant-density environment is expected in the core-collapse scenario when the wind from its progenitor is strong enough to overcome the deceleration ratio \citep[e.g., see][]{2017ApJ...848...15F,2019ApJ...879L..26F}. The analysis shows that for the synchrotron afterglow model, the stellar-wind scenario with ${\rm k=2}$ is the most favorable one, followed by ${\rm k=2.5}$, and for the SSC afterglow model, the stellar-wind scenario with ${\rm k=2.5}$ is the most favorable one followed by ${\rm k=2.0}$. The least favorable scenario is for constant-density medium in both the synchrotron and SSC afterglow models.\\

Irrespective of the circumburst medium, the synchrotron flux as a function of the density evolves as $F_\nu \propto A_{\rm k}^{0}\,(A_{\rm k}^{0})$ for ${\rm \nu_c^{syn} < \nu_{\rm LAT} < \nu_m^{syn} }$, $\propto A_{\rm k}^{\frac{10-p}{4(4-k)}}\,(A_{\rm k}^{\frac{2}{4-k}})$ for  ${\rm \nu_m^{syn} < \nu_{\rm LAT} < \nu_c^{syn} }$ and $\propto A_{\rm k}^{\frac{2-p}{4(4-k)}}\,(A_{\rm k}^{0})$ for ${\rm max\{\nu_m^{syn},\nu_c^{syn} \} < \nu_{\rm LAT}}$ for $1<p<2$ ($2<p$).   Similarly, the SSC flux  as a function of the density varies as $F_\nu \propto A_{\rm k}^{\frac{1}{2(4-k)}}\,(A_{\rm k}^{\frac{1}{2(4-k)}})$ for ${\rm \nu_c^{ssc} < \nu_{\rm LAT} < \nu_m^{ssc} }$, $\propto A_{\rm k}^{\frac{13-2p}{2(4-k)}}\,(A_{\rm k}^{\frac{11-p}{2(4-k)}})$ for  ${\rm \nu_m^{ssc} < \nu_{\rm LAT} < \nu_c^{ssc} }$ and $\propto A_{\rm k}^{\frac{2-p}{4-k}}\,(A_{\rm k}^{\frac{2-p}{2(4-k)}})$ for ${\rm max\{\nu_m^{ssc},\nu_c^{ssc} \} < \nu_{\rm LAT}}$ for $1<p<2$ ($2<p$).  It shows that an afterglow transition is expected in all cooling conditions for the SSC scenario but only in some cases for the synchrotron model. For instance, an afterglow transition is not expected when the synchrotron flux evolves in the condition ${\rm \nu_c^{syn} < \nu_{\rm LAT} < \nu_m^{syn}}$ for either $1<p<2$ or $2<p$. 

\subsection{The maximum energy photons of the synchrotron afterglow model}

During forward shocks, electrons are accelerated up to relativistic energies in a stratified environment through the Fermi mechanisms. The time required for an electron with Lorentz factor $\gamma_e$ to go from one side of the shock front to the other side is roughly equivalent to the Larmor time $t'_{\rm L}=\gamma_em_e c/(q_e B')$. It is possible to determine the maximum photon energy output by the synchrotron by equaling the accelerating ($t'_{\rm acc}=\xi 2\pi t'_{\rm L}$) and cooling ($t'_{\rm syn}=6\pi m_e c/(\sigma_T\gamma_e B'^2)$) time scales, with $\xi$ the Bohm parameter, and $\sigma_T$ the Thomson cross section. The limit for synchrotron photons  ($\nu^{\rm syn}_{\rm max}=\frac{q_e}{2\pi m_ec}(1+z)^{-1}\Gamma\gamma^{2}_{\rm max}B'$) for a uniform magnetic field becomes \citep{2012MNRAS.427L..40K}

\begin{equation}
\label{ene_max_q}
h\nu^{\rm syn}_{\rm max} \approx
2.7\times 10^2\, {\rm MeV}\left(\frac{1+z}{1.022}\right)^{\frac{k-5}{2(4-k)}}\,    A_{\rm k,-1}^{-\frac{1}{2(4-k)}}E_{53}^{\frac{1}{2(4-k)}}\,t_{2}^{-\frac{3-k}{2(4-k)}}\,.
\end{equation}


Within the 2FLGC, several bursts represent the extreme cases of Equation \ref{ene_max_q}. For instance, in the case of redshift variation, we have GRB 080916C and GRB 130702A with the redshifts $z=4.35$ and $z=0.14$, respectively. For a constant-density medium, and keeping all other parameters in Eq. \ref{ene_max_q} fixed, the maximum synchrotron energies would be MeV $h\nu^{\rm syn}_{\rm max,080916C}\approx96\ \rm MeV$ and $h\nu^{\rm syn}_{\rm max,130702A}\approx252\ \rm MeV$, while in a stellar-wind medium we would have $h\nu^{\rm syn}_{\rm max,080916C}\approx78$ and $h\nu^{\rm syn}_{\rm max,130702A}\approx249\ \rm MeV$. Therefore, the larger the redshift, the smaller the synchrotron energy and this effect is amplified more significantly with the value of stratification $k$. Similarly, in the case of LAT energy, the most significant variation in the catalog is between GRB 080916C ($E_{\rm \gamma, iso}=2.3\times10^{54}\ \rm erg$) and GRB 091127 ($E_{\rm \gamma, iso}=3\times10^{50}\ \rm erg$). In an analogous manner as before, in a constant-density (stellar-wind) medium we would have the maximum synchrotron energy $h\nu^{\rm syn}_{\rm max,080916C}\approx400\ \rm MeV$ ($590\ \rm MeV$) and $h\nu^{\rm syn}_{\rm max,091127}\approx130\ \rm MeV$ ($63\ \rm MeV$). As such, the synchrotron energy grows as the GRB's energy increases, as expected. Finally, in the case of time, we consider GRB 130427A with $t= 3.44\times 10^4\ \rm s$. In a constant-density (stellar-wind) medium, we obtain the maximum synchrotron energies $h\nu^{\rm syn}_{\rm max,130427A}\approx30\ \rm MeV$ ($63\ \rm MeV$), so we notice that at later times the maximum energy decreases. Considering this scenario and taking into account that the majority of bursts reported in the 2FLGC have redshifts with $z>1$, isotropic energies $E_{\rm \gamma, iso}\sim 10^{52}\,{\rm erg}$,  circumburst densities around $n (A_{\rm W})\simeq 1\, cm^{-3}\,(1)$ and durations larger than 100 s,  the limit for synchrotron photons is close or even less than 300 MeV. In this case, the LAT energy band could be almost described by the SSC process.

On the other hand, due to a lack of knowledge on how long it takes electrons to move from one side of the shock front to the other, this estimate might be off by a factor of 5-10 \citep[for a discussion, see][]{2001MNRAS.328..393A, 2006MNRAS.366..635L, 2007Natur.449..576U, 2012ApJ...749...80S}.\\

Synchrotron photons with energies significantly higher than that shown in Eq. (\ref{ene_max_q}) are possible under certain situations.  This requires the magnetic field to decrease behind the shock front on a scale far smaller than the path taken by electrons with the highest energies before they lose 50 percent of their original energy. In this case the magnetic field decays as $B(x)=B_{\rm w}(x/L_p)^{-\eta}+ B_0$, were $B_{\rm 0}$ and $B_{\rm w}$ are the weakest and strongest strengths of magnetic fields, respectively, and $L_p$ refers to the length-scale in which the field gradually loses its strength with $L_p\leq x$ \citep{2012MNRAS.427L..40K}.   The rate at which an electron loses energy to synchrotron radiation as it travels behind the shock front is

\be
\frac{d(\gamma_e m_e c^2)}{dt}=-\frac{\gamma_e\sigma_T c}{6\pi}\left[ B_{\rm w}\left(\frac{x}{L_p}\right)^{-\eta}+ B_0\right]^2\,.
\ee

The situation that attracts our attention is when electrons experience energy loss when they pass through an area of weak magnetic field. In this case, the limit for synchrotron photons becomes \citep{2012MNRAS.427L..40K}

\begin{equation}
\label{ene_max_q_mod}
h\nu^{\rm syn}_{\rm max} \approx
2.7\times 10^2\, {\rm MeV}\left(\frac{1+z}{1.022}\right)^{\frac{k-5}{2(4-k)}}\,    A_{\rm k,-1}^{-\frac{1}{2(4-k)}}E_{53}^{\frac{1}{2(4-k)}}\,t_{2}^{-\frac{3-k}{2(4-k)}}\left( \frac{B_{\rm w}}{B_{\rm 0}} \right)\,,
\end{equation}

which exceeds the value predicted by Eq. (\ref{ene_max_q}) by a factor of $B_{\rm w}/B_{\rm 0}$. Therefore, considering the magnetic field of the interstellar medium with $B_{\rm 0}=B_{\rm ism,-5}$ and $B_{\rm w}\simeq 8.3\,{\rm G}\,\left(\frac{1+z}{1.022}\right)^{\frac{3}{2(4-k)}} \epsilon^{\frac12}_{\rm B} A_{\rm k,-1}^{\frac{3}{2(4-k)}} E^{\frac{1-k}{2(4-k)}}_{53}t^{-\frac{3}{2(4-k)}}$, the maximum energy of synchrotron radiation becomes $h\nu^{\rm syn}_{\rm max}\sim 10^2\,{\rm TeV}$. In this scenario, the synchrotron afterglow model would almost describe the LAT energy band.\\

The synchrotron forward-shock model can only explain gamma-ray photons below the synchrotron limit as reported in Table 7 in the  2FLGC, which listed 29 bursts with photons exceeding energies $> 10\,{\rm GeV}$. On the other hand, since the IceCube team has reported no coincidences of neutrinos with GRBs \citep{2012Natur.484..351A, 2016ApJ...824..115A, 2015ApJ...805L...5A}, this data severely limits the number of hadrons involved in photo-hadronic interactions, which might yield photons with energy exceeding the synchrotron limit.   Given that the maximum photon energy generated by the SSC forward-shock model is above the LAT energy range ($h\nu^{\rm ssc}_{\rm max}=\gamma^2_{\rm max} h\nu^{\rm syn}_{\rm max}\gg 300\,{\rm GeV}$), the SSC mechanism is the most natural process to interpret the photons observed by the \textit{Fermi}-LAT instrument which is beyond the synchrotron limit. It is worth noting that although the maximum synchrotron energy for protons becomes $\sim 400\,{\rm GeV}$, this process is very ineffective, and it is pretty unlikely that it will play an essential role in GRBs.



\subsection{Evolution of the magnetic microphysical parameter}

The value of the magnetic microphysical parameter can be estimated through  Eq. \ref{nums_ssc} and equaling the synchrotron cooling break with the \textit{Fermi}-LAT band ($\nu^{\rm ssc}_{\rm c}=\nu_{\rm LAT}$).   This parameter corresponds to 

{\small
\begin{equation}
\label{eps_B2}
\varepsilon_B \lesssim
3.4\times10^{-5}\left(\frac{1+z}{1.022}\right)^{-\frac{3(k+2)}{7(4-k)}}\, (1+Y(\gamma_c))^{-\frac87}  A_{\rm k,-1}^{-\frac{18}{7(4-k)}}E_{53}^{\frac{7k-10}{7(4-k)}}t_{2}^{\frac{5k+7b(4-k)-2}{7(4-k)}} \left(\frac{h\nu^{\rm ssc}_c}{100\,{\rm MeV}}\right)^{-\frac27}\,, 
\end{equation}
}
which is in the usual range of those values ($10^{-5}\leq\varepsilon_B$) reported by modelling the multiwavelength GRB observations \citep{1999ApJ...523..177W, 2002ApJ...571..779P, 2003ApJ...597..459Y, 2005MNRAS.362..921P, 2014ApJ...785...29S}.

In the SSC processes, scattering photons by the relativistic electron can enter the  Klein - Nishina (KN) scattering regime. Above this critical energy, the SSC flux begins to be attenuated due to the quantum effects. Since the KN effects are not negligible, the value of the Compton parameter varies, and therefore it can be estimated as {\small $\frac{Y(\gamma_c)[Y(\gamma_c)+1]}{Y_N} =\left( \frac{\nu^{\rm syn}_{\rm m}}{\nu^{\rm syn}_{\rm c}} \right)^{-\frac{p-3}{2}}\,\left( \frac{\nu^{\rm syn}_{\rm KN}(\gamma_{\rm c})}{\nu^{\rm syn}_{\rm m}}\right)^\frac43 \,\,{\rm for}\,\,{\nu^{\rm syn}_{\rm KN}(\gamma_{\rm c}) <\nu^{\rm syn}_{\rm m} }$},  {\small $ \left(\frac{\nu^{\rm syn}_{\rm KN}(\gamma_{\rm c})}{\nu^{\rm syn}_c}\right)^{-\frac{p-3}{2}}\,\,{\rm for}\,\,    \nu^{\rm syn}_{\rm m} < \nu^{\rm syn}_{\rm KN}(\gamma_{\rm c}) < \nu^{\rm syn}_{\rm c}$},  {\small $1\,\, {\rm for} \,\, \nu^{\rm syn}_{\rm c} < \nu^{\rm syn}_{\rm KN}(\gamma_{\rm c})$} where {\small $Y_N=\frac{\varepsilon_{e}}{\varepsilon_{B}}  \left(\frac{\gamma_{\rm c}}{\gamma_{\rm m}}\right)^{2-p}$}  where $h\nu^{\rm syn}_{\rm KN} (\gamma_c) \simeq\frac{2\Gamma}{(1+z)}\,\frac{m_e c^2}{\gamma_c}$ for $\nu^{\rm syn}_{\rm m} < \nu^{\rm syn}_{\rm c}$ \citep[see][]{2009ApJ...703..675N, 2010ApJ...712.1232W}.  In addition, it is imperative to estimate the Lorentz factor of those electrons ($\gamma_{*}$) that could emit gamma-ray photons in the \textit{Fermi}-LAT energy range via the synchrotron process. In this case, a new spectral break $h\nu^{\rm syn}_{\rm KN}(\gamma_*)$ must be included in the SSC spectrum with a new Compton parameter $Y(\gamma_{*})$ given by {\small $Y(\gamma_*)=Y(\gamma_c) \left(\frac{\nu_{*}}{\nu_c}\right)^{\frac{p-3}{4}}   \left(\frac{\nu^{\rm syn}_{\rm KN}(\gamma_{\rm c})}{\nu^{\rm syn}_c}\right)^{-\frac{p-3}{2}}$} for $ \nu^{\rm syn}_{\rm m} <  h\nu^{\rm syn}_{\rm KN}(\gamma_*)=100\,{\rm MeV} < \nu^{\rm syn}_{\rm c} < \nu^{\rm syn}_{\rm KN}(\gamma_{\rm c})$  \citep[for details, see][]{2010ApJ...712.1232W}.  We can notice that for $Y(\gamma_c)\ll 1$, Eq. \ref{eps_B2} is not modified, and  for the case of $Y(\gamma_c)\gg 1$, the parameter $\epsilon_B$ increases. As a particular case, given the spectral breaks in the hierachy $ \nu^{\rm syn}_{\rm m} <  h\nu^{\rm syn}_{\rm KN}(\gamma_*) < \nu^{\rm syn}_{\rm c} < \nu^{\rm syn}_{\rm KN}(\gamma_{\rm c})$ and with $Y(\gamma_c)\gg 1$, the new Compton parameter becomes $Y(\gamma_*)\simeq 1$ which is ample used for modeling  multiwavelength afterglow observations \citep[see ][]{2010ApJ...712.1232W}. In each case, the parameter $\epsilon_B$ ranges in those values considered in \cite{2014ApJ...785...29S}.\\

Using the temporal and spectral indices, \cite{2019ApJ...883..134T} methodically examined the CRs of 59 LAT-detected bursts. The authors showed that these bursts satisfy the CRs of the slow-cooling regime, but only with the condition of having an exceptionally small value of $\epsilon_B<10^{-7}$. We found that in the SSC forward-shock scenario, the magnetic parameter ranges of $10^{-5}\lesssim\epsilon_B \lesssim 10^{-1}$, similar to those found after modeling  multiwavelength afterglow observations \citep[e.g., see][]{2014ApJ...785...29S}.

\subsection{Particular Burst: GRB 160509A}

\subsubsection{Multiwavelength Observations}

Both instruments of the \textit{Fermi} satellite detected GRB 160509A, GBM (Gamma Burst Monitor) and LAT, at 8:59:04.36 UTC on 2016-05-09 \citep{2016GCN.19411....1R, 2016GCN.19403....1L}. A radius of error of $0.50^\circ$ (90 percent containment, the systematic error alone) was used to pinpoint this burst, which was found at coordinates of ${\rm R.A.}= 310.1$ and ${\rm Dec}=76.0$ (J2000). At about $\approx77$~s s after the trigger, the LAT sensor observed an extremely intense photon with an energy of 52 GeV. Multiple peaks lasting $T_{90}=371$ s  were observed in the GBM light curve in the 50-300 keV energy range. On May 10 at 13:15 U.T., the GMOS-N and NIRI instruments at the Gemini North telescope detected optical spectroscopy and near-IR imaging, respectively, of this burst. The spectroscopic study revealed a redshift of z=1.17 \citep{2016GCN.19419....1T} based on a single, well-defined emission line ([OII] 3727 \AA). The VLA (Very Large Array) radio telescope picked up on this burst 0.36 days after the trigger time, in frequencies between 1.3 and 37 GHz \citep{2016ApJ...833...88L}.

\cite{2019ApJ...878...52A} derived the \textit{Fermi}-LAT light curve between  100 MeV and 100 GeV band of GRB 160905A, exhibiting two components a short-lasting peak at the end of the prompt episode and a long-lasting emission with a temporal break at 300 s. In addition,   \cite{2017ApJ...844L...7T} derived and analyzed the \textit{Fermi}-LAT spectrum, photon flux, and the evolution of the photon index between 0.1 and 1 GeV and 1 - 100 GeV bands of GRB 160905A. In particular, they reported photon emission in the 1 - 100 GeV band and no emission in the lower band (0.1 and 1 GeV) when the photon index was very hard ($\Gamma_{\rm LAT}=1.4\pm 0.3$).    \citet{2020ApJ...905..112F} modelled the \textit{Fermi}-LAT and \textit{Swift}-XRT observations of GRB 160509A. The GeV peak was described with the SSC emission from the reverse shock in the thick-shell regime, in a uniform-density medium, and the long-lasting component with synchrotron forward-shock emission. They showed that the passage of the forward shock synchrotron cooling break through the LAT band could explain the break in the long-lasting emission, and a fast evolution of cooling frequency could also explain the \textit{Swift}-XRT observations. The authors proposed that one of the possible scenarios with the potential to describe the fast evolution of the synchrotron cooling break and the plateau in the X-ray band is the variation of the microphysical parameters with $\epsilon_{\rm e}\propto t^{\rm 0.319^{+0.366}_{-0.375}}$ and $\epsilon_{\rm B} \propto t^{-\rm 1.401^{+0.136}_{-0.142}}$ in a very low circumburst density ($4.554^{+1.128}_{-1.121}\times 10^{-4}\,{\rm cm^{-3}}$).\\ 

\begin{figure*}
{\centering
\includegraphics[scale=1]{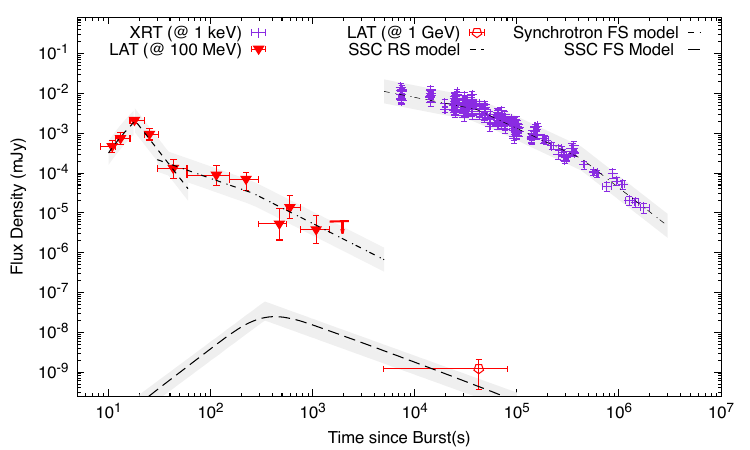}
\caption{\textit{Fermi}-LAT and \textit{Swift}-XRT observations of GRB 160509A with the best-fit curves generated by synchrotron forward-shock, and SSC reverse-shock afterglow model adapted from \cite{2020ApJ...905..112F}. In addition, we include the SSC flux evolving in constant-density medium using the best-fit values reported and the LAT data point at $4.2\times 10^4\,{\rm s}$ taken from \citet{2017ApJ...844L...7T}.} \label{Fig5}
}
\end{figure*}

Here, we estimate SSC flux in the scenario of the microphysical parameter variations using the best-fit values found by \cite{2020ApJ...905..112F} in order to describe the latest LAT data point at $4.2\times 10^4\,{\rm s}$ reported from \citet{2017ApJ...844L...7T}.   Figure \ref{Fig5} shows the 
\textit{Fermi}-LAT and \textit{Swift}-XRT observations of GRB 160509A with the best-fit curves generated by a synchrotron and SSC from forward and reverse shock, respectively, adapted from \cite{2020ApJ...905..112F}. The maximum photons energy radiated by synchrotron afterglow model is $h\nu^{\rm syn}_{\rm max} = 95.2\, {\rm MeV}$ at $10^4\,{\rm s}$.  While synchrotron flux cannot describe the latest LAT data point, SSC could marginally describe it. We conclude that a forward-shock scenario with microphysical parameter variations is needed for describing the early multi-wavelength observation of GRB 160509A. 

Although \cite{2016ApJ...833...88L} favored a uniform-density environment for this burst on physical grounds (based on the inferred initial Lorentz factor from the RS emission), they could not conclusively distinguish between a wind and a uniform-density medium based on multiwavelength observations alone. Although \cite{2020ApJ...905..112F} reported that a uniform-density medium is favored on the stellar-wind,  as shown here,  the analysis of the latest LAT observations confirmed the evolution of the forward shocks in a uniform-density environment.

\section{Closure relations of bursts with very-high-energy emission}\label{sec4}

\subsection{A GRB sample detected in VHEs with known CRs}


\subsubsection{GRB 180720B}
GRB 180720B was detected by \textit{Fermi}-GBM on July 20th 2018 at $T_0=$14:21:39.65 Universal Time (UT) \citep{Roberts_2018GCN.22981....1B} and five seconds after by \textit{Swift}-BAT \citep{Siegel_2018GCN.22981....1B}. Based on the duration of the prompt emission $T_{90} \sim 48.9 \pm 0.4$ sec, this burst was categorized as a long GRB. In addition, the isotropic energy released in the energy band of 50-300 keV was $E_{\rm \gamma, iso} = (6.0 \pm 0.1) \times 10^{53} \, \rm erg$. Furthermore, this GRB was also detected by \textit{Fermi}-LAT between $T_0$ and $T_0$+700 sec, with the most energetic photon of 5 GeV at $T_0$ + 137 sec \citep{Bissaldi_2018GCN.22980....1B}. The prompt emission detected by GBM was identified as extremely bright and in the top seven brightest.
Regarding the afterglow, the observations performed by the H.E.S.S. collaboration of VHE gamma-ray emission began 10.1 hours after $T_0$. It lasted for two hours  \citep{2019Natur.575..464A} revealing a gamma-ray excess associated with a point-like source centered at $RA=00^{\rm h} 0.2^{\rm min} 0.7.6^{\rm sec}$ and $Dec = 0.2^\circ 56' 06''$ (J2000). Similarly,  \textit{Swift}-XRT observed a bright afterglow 11 hours after $T_0$ in the 0.3-10 keV band, which was detectable even 30 days after $T_0$. The redshift was estimated to be $z=0.654$ by identifying several absorption features, such as Fe II, Mg II, Mg I, and Ca II \citep{Vreeswijk_2018GCN.22996....1V}.  

\subsubsection{GRB 190114C}
GRB 190114C was first detected by \textit{Swift}-BAT and \textit{Fermi}-GBM on 2019 January 14, $T_0=$20:57:03 UT \citep{Gropp_2019GCN.23688....1G, Hamburg_2019GCN.23707....1B}, categorising it as a long-GRB type with a $T_{90}$ of $\sim 116$ sec and $\sim 362$ sec by \textit{Swift}-BAT and \textit{Fermi}-GBM, respectively. The redshift estimation was $z=0.4245 \pm 0.0005$ \citep{2019GCN.23708....1C}. The isotropic equivalent energy released in the 10-1000 keV energy band during the prompt episode was $E_{\rm \gamma, iso} = 2.4 \pm 0.05 \times 10^{53} \, \rm erg$.
Approximately 57 seconds after the trigger alert, the MAGIC telescopes observed this burst until 15,912 seconds after the alert ($\sim 4.42$ hours), detecting VHE gamma-rays in the energy range of 0.2-1 TeV with more than 50 standard deviations in the first 20 minutes \citep{MAGIC_2019Natur.575..455M}. The afterglow multi-frequency observations from $5\times10^{-6}$ to $10^{12}$ eV were reported by \cite{MAGIC_2019Natur.575..459M}, who accounted a double-peaked feature in the broadband SED.

\subsubsection{GRB 190829A}
The \textit{Fermi}-GBM detected GRB 190829A on 29 August 2019 at $T_0=$19:55:53 UT, which was categorized as a long GRB class \citep{2019GCN.25551....1F}. After 51.6 sec, \textit{Swift}-BAT triggered and located this burst, and 97.3 sec after the BAT trigger, \textit{Swift}-XRT began its observation. The redshift was $z=0.0785$, confirmed by optical afterglow spectroscopy observations \citep{2019GCN.25565....1F}, turning this burst into one of the closest detected.    H.E.S.S. telescopes reported the detection of VHE gamma-rays with statistical significance of 21.7$\sigma$ and 5.5$\sigma$ during the first two nights. Their observations started 4.3 hours after the trigger time and ended 55.9 hours after the trigger, reporting the detection of VHE gamma-rays (from 0.18 to 3.3 TeV) during each night  \citep{HESS_2021Sci...372.1081H}.  The Fermi-LAT instrument observed the position of GRB 190829A \citep{2019GCN.25574....1P}, but no high-energy photons above 100 MeV were detected. The afterglow observations by the \textit{Swift} Ultraviolet/optical telescope began 158 seconds after the trigger time.

\subsubsection{GRB 221009A}
The \textit{Fermi}-GBM detected GRB 220019A on 2022 October 9 at $T_0=$13:16:59:99 UTC. It was categorized as a long GRB with a prompt emission that lasted more than 600 s \citep{Lesage_2023arXiv230314172L}. This GRB presented the highest total isotropic equivalent energy $E_{\gamma, \rm iso} = 1.0 \times 10^{53} \, \rm erg$. The estimation of the redshift found a value of $z=0.151$. \textit{Swift}-BAT began the observations of this burst 53 minutes and 17 seconds after the trigger time, yielding the best location at $RA=19^{\rm h} 13^{\rm min} 3.48^{\rm sec}$ and $Dec = 19^\circ 46' 24.6''$ (J2000). The Swift - XRT instrument began the observations 143 seconds after the trigger time \citep{Dichiara_2022GCN.32632....1D}. The most energetic \textit{Fermi}-LAT photon ever observed, with energy 99.3 GeV, arrived 240 seconds after the trigger time \citep{Pillera_2022GCN.32658....1P}. This burst was observed by LHAASO within 2000 seconds after the trigger time, detecting more than 5000 photons at VHE energies from 0.1 TeV extending up to 18 TeV \citep{Huang_2022GCN.32677....1H}.\\

\subsection{Analysis and Discussion}

Table \ref{tab:sample_vhe} shows the evolution of the temporal and spectral indexes reported by \textit{Fermi}-LAT \citep{2019Natur.575..459A, 2019ApJ...883..162F, 2019ApJ...885...29F, 2021ApJ...918...12F, Pillera_2022GCN.32658....1P} and TeV gamma-ray telescopes/observatories of a sample of 4 bursts (GRB 180720B, 190114C, 190829A and 221009A) with energetic photons above 100 GeV. These luminous bursts were detected by MAGIC \citep{2019Natur.575..459A, 2021ApJ...908...90A}, H.E.S.S. \citep{2019Natur.575..464A, HESS_2021Sci...372.1081H} and LHAASO \citep{Huang_2022GCN.32677....1H, 2023Sci...380.1390L}.

\begin{figure*}
    \centering
    \includegraphics[width=1.\textwidth]{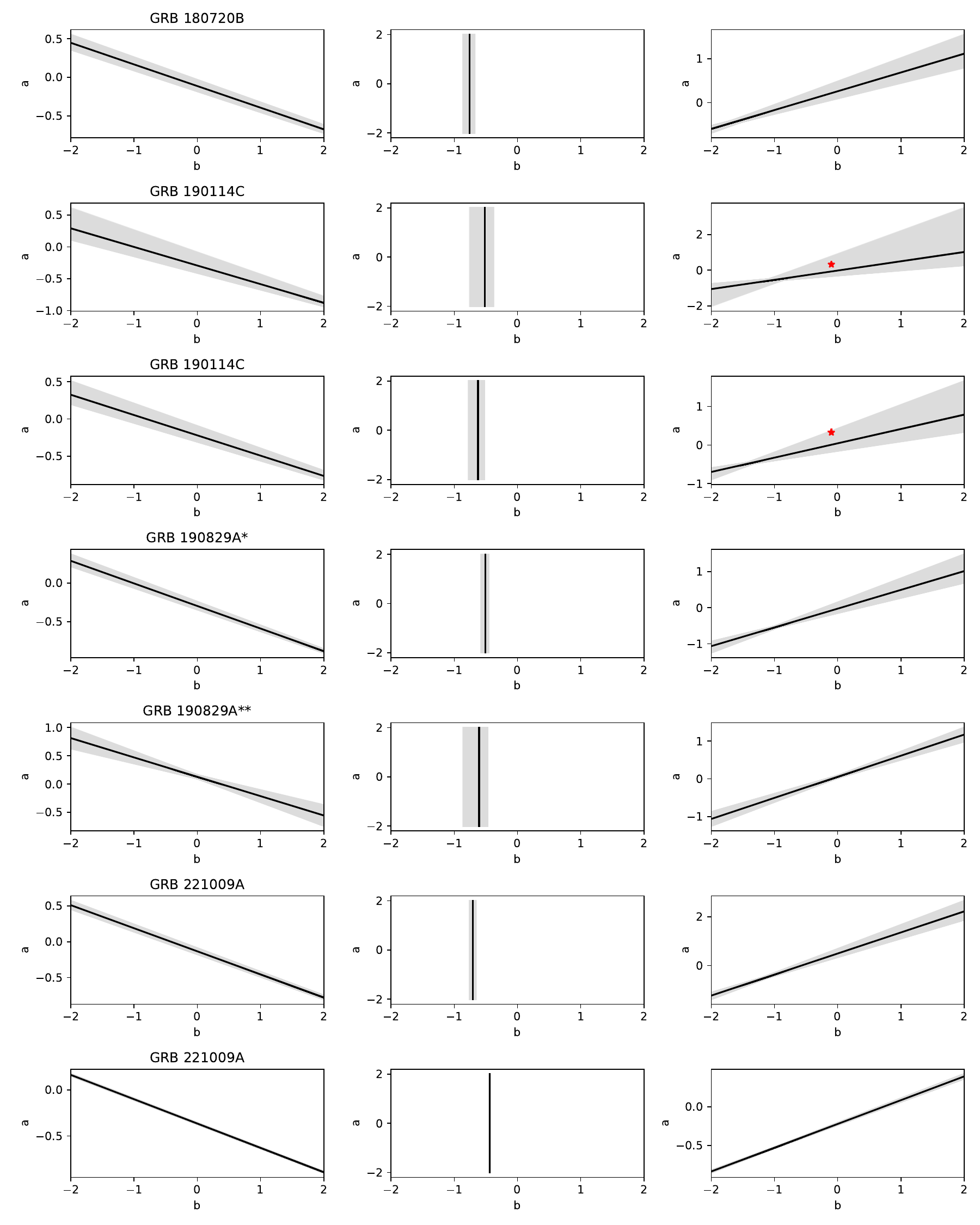}
    \caption{PL indices of microphysical variations for which the CRs are fulfilled for a constant-density medium.  The columns, from left to right, are in accordance with the cooling condition ${\rm max\{\nu_m^{ssc},\nu_c^{ssc} \} < \nu_{\rm LAT}}$, ${\rm \nu_c^{ssc} < \nu_{\rm LAT} < \nu_m^{ssc} }$ and  ${\rm \nu_m^{ssc} < \nu_{\rm LAT} < \nu_c^{ssc} }$, respectively. The small starlike symbols ($*$) and (${**}$) in GRB 190829A correspond to the first and second nights of observations.
}
    \label{Fig10}
\end{figure*}

\noindent Figure \ref{Fig10} presents the PL indices of the microphysical parameters for which the CRs are fulfilled for our GRB sample assuming a constant-density model. Each column corresponds to a different cooling condition; from left to right we have the cooling condition ${\rm max\{\nu_m^{ssc},\nu_c^{ssc} \} < \nu_{\rm LAT}}$, ${\rm \nu_c^{ssc} < \nu_{\rm LAT} < \nu_m^{ssc} }$ and  ${\rm \nu_m^{ssc} < \nu_{\rm LAT} < \nu_c^{ssc} }$, respectively. The first row refers to GRB 180720B, the second to the \textit{Fermi}-LAT observations of GRB 190114C, the third to the VHE observations of GRB 190114C, the fourth and fifth rows correspond to the first and second night of VHE observations of GRB 190829A, respectively. Finally, the sixth row pertains to the \textit{Fermi}-LAT observations of GRB 221009A, and the seventh row to the VHE observations of this GRB.   In general, we notice that for all GRBs, the relation between the power indices of the microphysical parameters depends on the cooling condition. However, each cooling condition leads to the same relation, as we notice by comparing each column. The condition ${\rm \nu_m^{ssc} < \nu_{\rm LAT} < \nu_c^{ssc} }$ shows an inversely proportional relation between $a$ and $b$, ${\rm max\{\nu_m^{ssc},\nu_c^{ssc} \} < \nu_{\rm LAT}}$ presents a proportional relation and the middle column, corresponding to ${\rm \nu_c^{ssc} < \nu_{\rm LAT} < \nu_m^{ssc} }$ predicts a negative value of $b$, but leaves $a$ as a free parameter.   Comparing the relations due to the \textit{Fermi}-LAT observations with those from the VHE observations, we notice that the lower-energy regime prefers slightly larger values of $a$. For instance, in the case of GRB 221009A, we see that the relations obtained for \textit{Fermi}-LAT are inconsistent with those for VHEs.

\begin{figure*}
    \centering
    \includegraphics[width=1.\textwidth]{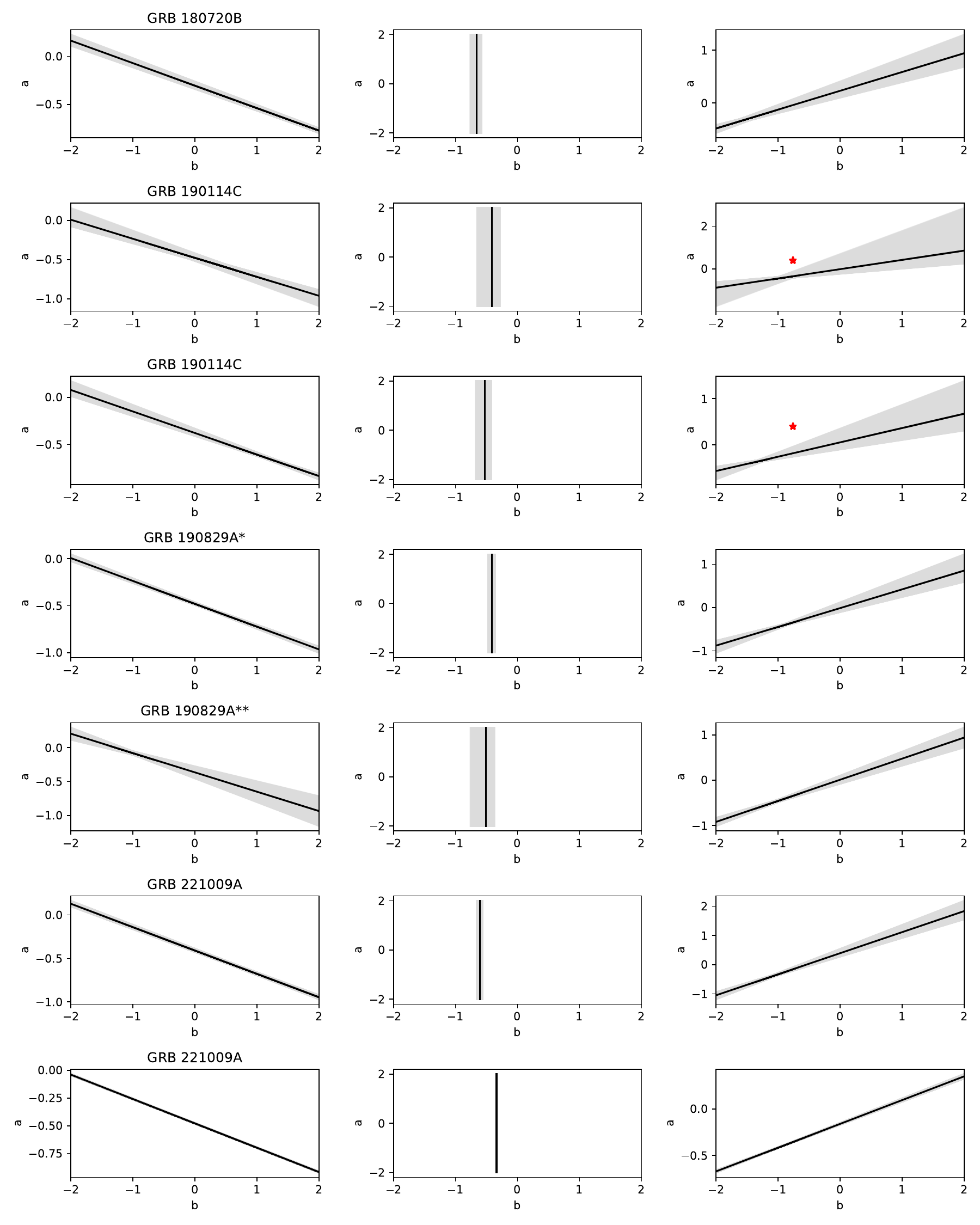}
    \caption{The same as Figure \ref{Fig10}, but for stellar-wind afterglow model.}
    \label{Fig11}
\end{figure*}

Figure \ref{Fig11} presents the PL indices of the microphysical parameters for which the CRs are fulfilled for the four GRBs in this section, assuming a stellar-wind density model. The layout is identical to that of Figure \ref{Fig10}. The same conclusions with the previous Figure apply to this one. The main difference between the two is found in the first column. We see that for the $k=2$ model, smaller values of $a$ are predicted. For the second column, we also note that the wind model predicts slightly smaller values of $b$. On the other hand, the third column is the same between both Figures, as the cooling condition ${\rm max\{\nu_m^{ssc},\nu_c^{ssc} \} < \nu_{\rm LAT}}$ does not depend on $k$.

We want to emphasize that the no association of neutrinos with these powerful GRBs suggests that the ranges of values of $a>0$ and $b<0$ in Figures \ref{Fig10} and \ref{Fig11} are favored. It occurs when the stored magnetic field energy is released by additional magnetic reconnections, as predicted in some dissipation models, such as the ICMART scenario \citep{2011ApJ...726...90Z}.


\subsubsection{GRB 180720B} 
 Table \ref{tab:sample_vhe} shows that the H.E.S.S. collaboration did not report the temporal index of the VHE gamma-ray emission \citep{2019Natur.575..464A}, so we only analyze the CRs reported by Fermi-LAT. The spectral analysis performed by LAT instrument led to an spectral index of $\beta_{\rm LAT}=1.15\pm 0.10$. Taking into account the CRs of the standard SSC afterglow model ($a=0$ and $b=0$) in the cooling condition (${\rm max\{\nu_m^{ssc},\nu_c^{ssc} \} < \nu_{\rm LAT}}$), the theoretical values of temporal indexes are $\alpha=1.34\pm 0.23$ and $1.30\pm 0.20$ for a homogeneous and stellar-wind environment, respectively, which are inconsistent with the value of temporal index reported by MAGIC ($\alpha_{\rm LAT}=1.90\pm0.10$). It is worth nothing that other cooling condition would lead to atypical values of the spectral index of electrons (e.g. $p>3$).  A possible explanation is to consider the variation of the microphysical parameters during the deceleration phase. In this case,  the microphysical parameters would vary following the relation $5.2a-1.85b=1.125$ for ${\rm k=0}$, and  $5.2a-1.85b=1.2$ for ${\rm k=2}$.

 On the other hand, we discuss the top panels in Figures \ref{Fig10} and \ref{Fig11} around the scenario of the decaying microturbulence.  For the cooling condition ${\rm max\{\nu_m^{ssc},\nu_c^{ssc} \} < \nu_{\rm LAT}}$, the parameter $a$ lies in range of $-0.08\leq a \leq -0.39$ for the scenario of gradually decaying microturbulence and   $a\leq -0.39$ for fast decaying microturbulence in both stellar-wind and ISM.  For the cooling condition ${\rm \nu_m^{ssc} < \nu_{\rm LAT} < \nu_c^{ssc} }$, the parameter $a$  lies in the range of $0.15\leq a \leq 0.82$ ($0.12\leq a \leq 0.76$) for gradually decaying microturbulence and $a\leq 0.15$ ($a\leq 0.12$) for fast decaying microturbulence in constant-density medium (wind).

\subsubsection{GRB 190114C} 

 \cite{2021MNRAS.504.5685M} analyzed the radio and optical afterglow observations of TeV-bright GRB 190114C in the range of $\sim 1$ to $\sim 140$ days.  They found that these observations did not follow the CRs of the standard synchrotron forward-shock model. Therefore, they required that the microphysical parameters evolved in time, as $\epsilon_{\rm e}\propto t^{-0.4}$ and $\epsilon_{\rm B}\propto t^{0.1}$ for ISM, and $\epsilon_{\rm e}\propto t^{-0.4}$ and $\epsilon_{\rm B}\propto t^{0.76}$ for stellar-wind. To compare which afterglow model was consistent with MAGIC observation's temporal and spectral index, we include these values with red stars in Figures \ref{Fig10} and \ref{Fig11} for constant-density and stellar-wind environments, respectively. These Figures show that a constant-density medium is self-consistent and not a stellar-wind medium. In this case, the SSC afterglow model with variation of microphysical parameters and evolving in constant-density medium is the afterglow scenario for describing GRB 190114C.

\paragraph{LAT CRs.}  For the cooling condition ${\rm max\{\nu_m^{ssc},\nu_c^{ssc} \} < \nu_{\rm LAT}}$, the parameter $a$ lies in range of $-0.65\leq a \leq -0.28$ for the scenario of gradually decaying microturbulence and   $a\leq -0.65$ for fast decaying microturbulence in both stellar-wind and ISM.  For the cooling condition ${\rm \nu_m^{ssc} < \nu_{\rm LAT} < \nu_c^{ssc}}$, the parameter $a$  lies in range of $-0.05\leq a \leq 0.32$ ($-0.02\leq a \leq 0.30$) for gradually decaying microturbulence  and $a\leq -0.05$ ($a\leq 0.02$) for fast decaying microturbulence in ISM (wind).

\paragraph{MAGIC CRs.} For the cooling condition ${\rm max\{\nu_m^{ssc},\nu_c^{ssc} \} < \nu_{\rm MAGIC}}$, the parameter $a$ lies in range of $-0.51\leq a \leq -0.19$ for the scenario of gradually decaying microturbulence and   $a\leq -0.19$ for fast decaying microturbulence in both stellar-wind and constant-density medium.  For the cooling condition ${\rm \nu_m^{ssc} < \nu_{\rm MAGIC} < \nu_c^{ssc} }$, the parameter $a$  lies in range of $-0.02\leq a \leq 0.2$ ($-0.01\leq a \leq 0.18$) for gradually decaying microturbulence  and $a\leq -0.02$ ($a\leq -0.01$) for fast decaying microturbulence in constant-density medium (wind).

\subsubsection{GRB 190829A}

The Fermi-LAT instrument did not detect photons with energies greater than 100 MeV. Still, the H.E.S.S. collaboration reported the detection of VHE gamma-rays during the first two nights of observation, as shown in Table \ref{tab:sample}. Therefore, we only analyze the CRs reported by H.E.S.S. around the first (*) and second (**) night. 

The H.E.S.S. team reported the detection of VHE gamma-rays with statistical significances of 21.7$\sigma$ and 5.5$\sigma$, during two consecutive nights \citep{HESS_2021Sci...372.1081H}. The spectral analysis in the energy range of 0.1 - 3.3 TeV resulted in photon indices of $\beta_{\rm HESS}=1.06\pm 0.10 $ and $0.86\pm 0.26 $ for the first and second nights of observations. A temporal analysis in the 0.2 - 3.0 TeV energy range with a PL function exhibited a temporal index of $\alpha_{\rm HESS}=1.09\pm 0.05$. A potential explanation for this behavior where the spectral index varies, and the temporal index remains constant could be given in terms of the SSC afterglow scenario with variation of the microphysical parameters. With the values of spectral indexes,  it can be observed that the spectral index of the second night is a little hard concerning the first one, whereas the temporal index remains constant. Given that the temporal index is the same for both nights, it is reasonable to assume that the SSC flux lies in the same cooling condition (${\rm max\{\nu_m^{ssc},\nu_c^{ssc} \} < \nu_{\rm HESS}}$); otherwise, unreasonable values are obtained.   Using the CRs of the standard SSC afterglow model ($a=0$ and $b=0$), the theoretical temporal indexes when the first and second night is considered are $\alpha=1.14\pm 0.20$ and $0.69\pm 0.52$  for a homogeneous medium, and $\alpha=1.12\pm 0.20$ and $0.72\pm 0.52$ for stellar-wind environment, respectively.  The observed temporal index ($\alpha_{\rm HESS}=1.09\pm 0.05$) strongly agrees with the standard SSC CRs for a homogeneous medium and stellar wind during the first night, but it does not with the ones of the second night. This could suggest that between both nights, the microphysical parameters vary following the relation $1.44a-1.07b = 0.455$ for $k=0$ and by $1.44a-1.07b=0.4$ for $k=2$. We could conclude that variations of spectral index keeping the temporal index constant are associated with the evolution of microphysical parameters.
 

\paragraph{First night of HESS observations.}

For the cooling condition ${\rm max\{\nu_m^{ssc},\nu_c^{ssc} \} < \nu_{\rm HESS}}$, the parameter $a$ lies in range $-0.68\leq a \leq -0.36$ for the scenario of gradually decaying microturbulence, and  $a\leq -0.68$ for fast decaying microturbulence in both stellar-wind and ISM.  For the cooling condition ${\rm \nu_m^{ssc} < \nu_{\rm HESS} < \nu_c^{ssc} }$, the parameter $a$  lies in the range $-0.02\leq a \leq 0.48$ ($-0.01 \leq a \leq 0.46$) for gradually decaying microturbulence  and $a\leq -0.02$ ($a\leq -0.01$) for fast decaying microturbulence in constant-density medium (wind).

\paragraph{Second night of HESS observations.}

For the cooling condition ${\rm max\{\nu_m^{ssc},\nu_c^{ssc} \} < \nu_{\rm HESS}}$, the parameter $a$ lies in range $-0.76\leq a \leq -0.39$ for the scenario of gradually decaying microturbulence, and  $a\leq -0.76$ for fast decaying microturbulence in both stellar-wind and ISM.  For the cooling condition ${\rm \nu_m^{ssc} < \nu_{\rm HESS} < \nu_c^{ssc} }$, the parameter $a$  lies in the range $-0.05\leq a \leq 0.52$ ($-0.04\leq a \leq 0.56$) for gradually decaying microturbulence  and $a\leq -0.05$ ($a\leq -0.04$) for fast decaying microturbulence in constant-density medium (wind).

\subsubsection{GRB 221009A}

The Water Cherenkov Detector Array (WCDA) instrument of LHAASO detected GRB 221009A with a statistical significance $> 250\sigma$. During the first $\sim 3000\,{\rm s}$ after the trigger time, WCDA reported $6.4\times 10^{4}$ photons in the energy range $\sim 0.2 - 7\,{\rm TeV}$. This observatory reported three temporal indexes: at the earliest times $\alpha_{\rm LHAASO, 1}=1.82^{+0.21}_{-0.18}$, at intermediate times $\alpha_{\rm LHAASO, 2}=-1.115^{+0.012}_{-0.012}$, and finally, at late times  $\alpha_{\rm LHAASO, 3}=-2.21^{+0.03}_{-0.83}$. On the other hand, the collaboration reported a single spectral index in the small range $2.324\pm 0.065 < \beta_{\rm LHAASO}< 2.429\pm 0.062$ during the entire temporal coverage \citep{2023Sci...380.1390L}. One possible rationale for the observed phenomenon, in which the temporal index exhibits variations but the spectral index remains constant, may be attributed to the scenario of SSC afterglow, where evolution in microphysical parameters occurs. While $\alpha_{\rm LHAASO, 2}$ and $\alpha_{\rm LHAASO, 3}$ can be associated to the deceleration and the post-jet break phases, respectively, of the standard SSC model evolving during the cooling condition (${\rm max\{\nu_m^{ssc},\nu_c^{ssc} \} < \nu_{\rm HESS}}$) in a constant-density medium or a stellar-wind \citep[see][]{2023Sci...380.1390L}, the initial temporal index of $\alpha_{\rm LHAASO, 1}=1.82^{+0.21}_{-0.18}$ could be consistent with the SSC afterglow scenario during the deceleration phase with evolution of microphysical parameters in both a constant-density medium or in a stellar-wind. For instance, in the homogeneous medium, the relation as a function of $a$ and $b$ is $4.8a - 1.9b + 6.09=0$, and in the stellar wind, the relation is $4.8a - 1.9b + 6.04=0$. Considering that a very early jet break has never been observed, the last temporal index of $\alpha_{\rm LHAASO, 3}=-2.21^{+0.03}_{-0.83}$ could also be explained in the SSC scenario evolving in a constant-density medium with the variation of microphysical parameters. In this case, the relation as a function of $a$ and $b$ is $4.8a - 1.9b - 1.97=0$. We could conclude, in general, that variations of the temporal index keeping the spectral index constant are associated with the evolution of microphysical parameters.

\paragraph{LAT CRs.}  For the cooling condition ${\rm max\{\nu_m^{ssc},\nu_c^{ssc} \} < \nu_{\rm LAT}}$, the parameter $a$ lies in the range $-0.79\leq a \leq -0.81$ for the scenario of gradually decaying microturbulence and   $a\leq -0.81$ for fast decaying microturbulence in both stellar-wind and ISM.  For the cooling condition ${\rm \nu_m^{ssc} < \nu_{\rm LAT} < \nu_c^{ssc} }$, the parameter $a$  lies in the range $0.15\leq a \leq 1.21$ ($0.12\leq a \leq 1.13$) for gradually decaying microturbulence  and $a\leq 0.15$ ($a\leq 0.12$) for fast decaying microturbulence in homogeneous medium (wind).

\paragraph{LHAASO CRs.}  For the cooling condition ${\rm max\{\nu_m^{ssc},\nu_c^{ssc} \} < \nu_{\rm LHAASO}}$, the parameter $a$ lies in range of $-0.79\leq a \leq -0.49$ for the scenario of gradually decaying microturbulence and   $a\leq -0.79$ for fast decaying microturbulence in both stellar-wind and ISM.  For the cooling condition ${\rm \nu_m^{ssc} < \nu_{\rm LHAASO} < \nu_c^{ssc} }$, the parameter $a$  lies in range of $-0.20\leq a \leq 0.14$ ($-0.19\leq a \leq 0.15$) for gradually decaying microturbulence  and $a\leq -0.20$ ($a\leq -0.19$) for fast decaying microturbulence in homogeneous medium (wind).


\section{Discussion: Previous works and our contribution}\label{sec5}

It is assumed that the microphysical parameters, the fraction of energy that goes into electron $\epsilon_e$ and magnetic fields $\epsilon_B$, do not vary in time. This stationary assumption leads to good consistency between models and observations of late afterglows.   However, this assumption has been challenged due to different reasons, including the description of the plateau phase in the X-ray observations \citep{2006A&A...458....7I}, the modeling of afterglow observations \citep{2003ApJ...597..459Y, 2006MNRAS.369.2059P}, the interpretation of fast evolution of the spectral breaks \citep{2020ApJ...905..112F},  the explanation of magnetic field amplification and microturbulence, which in turn guides the complexly non-linear acceleration process, among others \citep{2003MNRAS.339..881R, 2005PThPh.114.1317I, 2013arXiv1305.3689L,2013MNRAS.428..845L}.

\cite{2006A&A...458....7I} explored the time variation of the parameters to explain the plateau phase of many X-ray light curves and found that this behavior could be explained if $\epsilon_e\propto t^{-a}$ with $a=1/2$. Modeling of radio afterglow observations of GRB 970508 by \cite{1999ApJ...523..177W} at 12 days and by \cite{1998ApJ...497..288W} at $\sim$ 1 year led to different values of microphysical parameters; the former reported $\epsilon_e=0.12$ and $\epsilon_B=0.089$, and the later $\epsilon_e\simeq\epsilon_B=0.5$. \cite{2018ApJ...859..163H} modeled the early-time emission of GRB 120729A by considering a variation of the magnetic parameter given by a BPL of the form $\epsilon_B\propto t^{-\rm b}$ with $\mathrm{b}_{1}=-(0.18\pm0.04)$ and $\mathrm{b}_{2}=-(0.84\pm0.04)$.   \cite{2003ApJ...597..459Y} performed a study of four well-observed bursts: GRB 970508, GRB 980329, GRB 980703, and GRB 000926. The authors first fit the observations with a model with constant microphysical parameters and noted diversity in the values of $\epsilon_{\rm B}$, namely that $0.002<\epsilon_{\rm B}<0.25$. Afterward, they proposed the relation $\epsilon_{\rm B}\propto\Gamma^{\alpha_x}$ with $-2\leq \alpha_x\leq1$, which led to a difference by up to an order of magnitude from the previous model's parameters. However, the authors noted that the results were not unique, so conclusions on the values could not be made. In a similar fashion, \cite{10.1111/j.1365-2966.2009.15886.x} analyzed GRB 060206, GRB 070311 and GRB 071010A with a two-region model ($1,2$) in which the microphysical parameters varied as $\epsilon_{j}=\epsilon_{j,0}\Gamma^{-\alpha_i}$, where $j\in[\rm e,B]$ and $i\in[1,2]$. The authors were able to reproduce the observed R-band and X-ray light curves with their model and found that $\epsilon_{\rm e,0}=0.3$ and $\epsilon_{\rm B}=0.03$ were sufficient to fit all GRBs, but different values of $\alpha_i$ were required.   \cite{2006MNRAS.369.2059P} required the evolution of microphysical parameters as $\epsilon_{\rm B} \propto \Gamma^{-\alpha_b}$ and $\epsilon_{\rm e}\propto \Gamma^{-\alpha_e}$ to model the X-ray and optical afterglow observations of GRB 050319, GRB 050401, GRB 050607, GRB 050713A, GRB 050802 and GRB 050922C which exhibited a steepening at $\sim$ 1-4 h after the prompt episode without displaying a break in the optical emission. They found that indexes $\alpha_b$ and $\alpha_e$ were in the range of  $4.2\leq \alpha_b\leq 7.6$ and $-4.1\leq\alpha_e\leq1.7$.   \cite{2003ApJ...597..459Y}, \cite{2006MNRAS.369..197F} considered, among other options, a model for the variation of microphysical parameters of the form $\epsilon_{\rm e}\propto\Gamma^{-a}$ and $\epsilon_{\rm B}\propto\Gamma^{-b}$. They found that the X-ray afterglow observations of GRB 050319, GRB 050401, and GRB 050315 could be well explained with $0.6\leq a \leq 0.7$, and $0.45\leq b \leq 1.2$, which implied an increase in time of the microphysical parameters. The authors noted, however, that while the X-ray afterglow was well-fitted, this model could not properly account for the R-band afterglow.    \cite{2006MNRAS.370.1946G} briefly considered a model where $\epsilon_{\rm e}\propto t^{\alpha_{\rm e}}$ and $\epsilon_{\rm B}\propto t^{\alpha_{\rm B}}$ to explain an observed stage of flattish decay in \textit{Swift} X-ray afterglow light curves. The authors found that the relation between the power indices $\alpha_{\rm e}+\alpha_{\rm B}\sim1-2$, was required to reproduce these observations, which led them to consider linear growth in time of either or both of the microphysical parameters as a possible explanation.

Relativistic collisionless shock front formation in a weakly magnetized surrounding medium is explained by the self-generation of intense small-scale electromagnetic fields mediating the transition between the upstream unshocked and the downstream shocked media. This scenario was considered by \cite{2003MNRAS.339..881R}, \cite{2005PThPh.114.1317I}  and \cite{2013MNRAS.428..845L} that showed how the spectrum is modified if a short magnetic length-scale is considered.   \cite{2005PThPh.114.1317I} suggested that coherent radiation could happen in relativistic collisionless shocks via two-stream Weibel instabilities. They argued that the coherence amplifies the radiation power, and electrons/protons cool quickly before being randomized. \cite{2013MNRAS.428..845L} considered the variation of magnetic parameter $\epsilon_B\propto t^{-b}$ and discussed the effect of the parameter $b$ on the resulting spectrum for synchrotron and dominant inverse Compton, identifying two scenarios. In the first scenario, turbulence gradually decays with $-1<b<0$ and $b>-4/(p+1)$ for synchrotron and dominant inverse Compton, respectively, and in the second scenario, turbulence rapidly decays with $b<-1$ for synchrotron model and  $-3<b<-4/(p+1)$ for dominant inverse Compton.   \cite{2013arXiv1305.3689L} described the multiwavelength observations of GRB 090902B, GRB 090323, GRB 090328 and GRB 110731A, using a  synchrotron model with variation only of the magnetic microphysical parameter $\epsilon_B\propto t^{-b}$  with $b$ in range of $0.4\lesssim b \lesssim 0.5$. They associated this variation with the PL-decaying micro-turbulence and the magnetic field amplification, which could occur in the relativistic collisionless shock of  GRB afterglows.

\cite{2020ApJ...905..112F} modelled the \textit{Fermi}-LAT and \textit{Swift}-XRT observations of GRB 160509A.  They showed that the fast evolution of cooling frequency exhibited in the LAT and XRT instruments and the plateau in the X-ray band could be interpreted as the variation of the microphysical parameters $\epsilon_{\rm e}\propto t^{-\rm a}$ and $\epsilon_{\rm B} \propto t^{-b}$ with  $a= 0.319$ and $b= 1.401$. In a more recent article, \cite{2021MNRAS.504.5685M} analyzed the TeV-bright long gamma-ray burst 190114C and found that the microphysical parameters of the forward shock evolve in time. The authors considered two possibilities for the circumburst density profile ($k=0,2$) and concluded that time-dependence of both $\epsilon_e$ and $\epsilon_B$ is necessary to reproduce the observations in both scenarios. Furthermore, motivated by the analysis of the afterglow's SED in several energy bands, the researchers were able to fix the value of the electron spectral index to $p=2.01$ and the power of the $\varepsilon_e$ variation to $\mathrm{a}=0.4$. This, in turn, allowed them to analyze the $\mathrm{b}-k$ parameter space that was consistent with the observations, which they showed in Figure 9. In summary, they found that for $k=0$, $\mathrm{b}\in[-0.2,0.5]$ was allowed, while for $k=2$, $\mathrm{b}\in[-0.8,-0.6]$ was permitted.

Previous studies that considered the evolution of microphysical parameters have focused on the role of this variation as a means to describe particular segments of GRB light curves and, in a few cases, to explain magnetic field amplification and microturbulence during the shocks. Given the multiwavelength observations, they have also only addressed synchrotron afterglow models and considered, in all cases, an electron population with a PL index $2<p$, but not with a hard spectral index ($1\leq p \leq 2$). Authors have also considered differences between afterglow models with constant circumburst medium ($k=0$) and stellar-wind ($k=2$) but have not tested different types of stratification $n\propto r^{-k}$ with a general $k$. In the current work, we have derived and shown the SSC and synchrotron forward-shock model including the light curves and CRs with the variation of the microphysical parameters  ($\epsilon_{\rm e}\propto t^{-\rm a}$ and $\epsilon_{\rm B} \propto  t^{-\rm b}$) evolving in the stratified environment with a density profile $\propto r^{-k}$ with ${\rm k}$ in the range of  $0\leq k < 3$. In addition, we have considered the electron distribution in the entire range of spectral index $1\leq p \leq 2$ and $ 2 < p$.   We have applied the current model to investigate the evolution of the spectral and temporal indexes of the GRB sample reported in 2FLGC. Moreover, we have also considered a sub-sample of bursts from 2FLGC with spectral index $\beta<1$ and temporal index $\alpha>1.3$, which can hardly be modeled with the CRs of the standard synchrotron forward-shock model. In this case, a hard spectral index is favored. We have explicitly shown the evolution of the synchrotron and SSC light curves with microphysical parameter variations and demonstrated that a plateau phase or a MeV-GeV peak could be expected in the LAT energy range depending on the parameter values of $a$ and $b$. This is the case of GRB 090510, GRB 090902B, and GRB 160509A, which exhibited a plateau phase, and GRB 080916C, GRB 090510A, GRB 090902B, GRB 090926A, GRB 110731A, GRB 130427A and GRB 160509A that showed a MeV - GeV peak.  In addition, we found that the cooling conditions for synchrotron afterglow scenario ${\rm \nu_c^{\rm ssc} < \nu_{\rm LAT} < \nu_c^{\rm ssc}}$ and ${\rm max\{\nu_m^{\rm syn},\nu_c^{\rm syn} \} < \nu_{\rm LAT}}$ could exhibit plateaus for both a homogeneous medium and a stellar wind. It can occur when the stored magnetic field energy is released by additional magnetic reconnections, as predicted in some dissipation models, such as the ICMART scenario.  Similarly, the cooling conditions for the SSC model ${\rm \nu_c^{\rm ssc} < \nu_{\rm LAT} < \nu_c^{\rm ssc}}$ and ${\rm max\{\nu_m^{\rm syn},\nu_c^{\rm syn} \} < \nu_{\rm LAT}}$ could exhibit high-energy plateaus. In this case,  the variation of microphysical parameters could be associated with the continuous energy injection with different efficiencies.   We have plotted the SSC light curves and spectra with the CTA (Southern array), MAGIC, and Fermi-LAT sensitivities and show that using the exact quantities of kinetic energies, circumburst densities or spectral indexes, but different evolution of the microphysical parameters, this emission could be observed in CTA (Southern array), MAGIC and Fermi-LAT. We have applied the CRs of SSC and synchrotron afterglow model in each cooling condition to the entire sample of GRBs reported in the 2FLGC using MCMC simulations and a selection of 4 bursts (GRB 180720B, 190114C, 190829A and 221009A) with energetic photons above 100 GeV, which MAGIC, H.E.S.S. and LHAASO gamma-ray observatories detected.   We have found that the most likely afterglow model using synchrotron and SSC emission on the 2FLGC corresponds to the constant-density scenario. Given the best-fit values of the parameter $b$ from the 2FLGC, we have found that  only the cooling conditions ${\rm \nu_m^{\rm syn} < \nu_{\rm LAT} < \nu_c^{\rm syn} }$ and ${\rm \nu_m^{\rm ssc} < \nu_{\rm LAT} < \nu_c^{\rm ssc} }$ lie in the scenario of rapidly decaying microturbulence for both synchrotron and SSC processes.   Finally, we have found that variations of the spectral (temporal) index keeping the temporal (spectral) index constant are associated with the evolution of microphysical parameters.

\section{Conclusion}
\label{sec6}

We have derived the synchrotron and SSC forward-shock scenario, including the light curves and CR, with variations in the microphysical parameters
($\epsilon_{\rm e}\propto \varepsilon_{\rm e} t^{-\rm a}$ and $\epsilon_{\rm B} \propto \varepsilon_{\rm B} t^{-\rm b}$).  We have considered the forward shock evolves in a stratified environment with density-profile described by $n(r) \propto r^{\rm -k}$ with $k=0.0$, $0.5$ $1.0$, $1.5$, $2.0$ and $2.5$ and relativistic electron population with a spectral index of $1<p<2$ and $p>2$. We have performed an analysis between these CRs and the spectral and temporal indices of bursts that were reported in 2FLGC. Considering the synchrotron and SSC the light curves,  we have shown that a plateau phase or a MeV-GeV peak could be expected in the LAT energy range depending on the parameter evolution. Potential candidates for a plateau phase are GRB 090510, GRB 090902B, and GRB 160509A, and for a MeV - GeV peak, are GRB 080916C, GRB 090510A, GRB 090902B, GRB 090926A, GRB 110731A, GRB 130427A and GRB 160509A.   We have plotted the SSC light curves and spectra with the CTA (Southern array), MAGIC, and Fermi-LAT sensitivities and show that using the exact quantities of kinetic energies, circumburst densities or spectral indexes, but different evolution of the microphysical parameters, this emission could be observed in CTA (Southern array), MAGIC and Fermi-LAT. 

We have considered LAT-detected bursts reported in 2FLGC and applied the MCMC method to find the best-fit values that satisfy the CRs in each cooling condition for the fast- and slow-cooling regimes.  The number and percentage of bursts following each cooling condition for the synchrotron and SSC afterglow model evolving in a stratified medium are as follows.  In the synchrotron afterglow scenario,  for the cooling condition ${\rm \nu_m^{syn} < \nu_{\rm LAT} < \nu_c^{syn} }$ the best case is most consistent with ${\rm k=0, 1.0, 1.5, 2}$ (23 GRBs, 27.06\%). For the cooling condition ${\rm \nu_c^{syn} < \nu_{\rm LAT} < \nu_m^{syn} }$, all cases of ${\rm k}$ are equally likely (1 GRB, 1.18\%).  For the cooling condition ${\rm max\{\nu_m^{syn},\nu_c^{syn} \} < \nu_{\rm LAT}}$, the best case is most consistent with ${\rm k=1}$ (24 GRBs, 28.24\%) and worst scenario is for ${\rm k=2.5}$ (21 GRBs, 24.71\%).  In SSC afterglow scenario, for the cooling condition ${\rm \nu_m^{ssc} < \nu_{\rm LAT} < \nu_c^{ssc} }$ the best case is most consistent with ${\rm k=0}$ (23 GRBs, 27.06\%) followed by all cases with 22 GRBs (25.88\%).  For the cooling condition ${\rm \nu_c^{ssc} < \nu_{\rm LAT} < \nu_m^{ssc}}$, all cases of ${\rm k}$ are equally likely (1 GRB, 1.18\%).  For the cooling condition ${\rm max\{\nu_m^{ssc},\nu_c^{ssc} \} < \nu_{\rm LAT}}$, the best cases are most consistent with ${\rm k=0}$ and ${\rm k=0.5}$ (24 GRBs, 28.24\%).   In particular, we have also considered a sub-sample of bursts from 2FLGC with spectral index $\beta<1$ and temporal index $\alpha>1.3$, which can hardly be modeled with the CRs of the standard synchrotron forward-shock model. In this case, a hard spectral index is favored. \cite{2019ApJ...883..134T} meticulously analyzed the CRs of 59 LAT-detected bursts by looking at temporal and spectral indexes. For instance, they discovered that a sizable percentage of bursts could only be inadequately characterized using the standard synchrotron emission model, even though this model accurately describes the spectrum and temporal indices in most cases. Nonetheless, they demonstrated that many GRBs fulfill the CRs of the slow-cooling regime, but only when the magnetic microphysical parameter has a value of $\epsilon_B<10^{-7}$ or smaller. Here, we show that a value of $10^{-5}\lesssim\epsilon_B\lesssim 10^{-1}$ \citep[e.g., see][]{2014ApJ...785...29S} is generated by the SSC afterglow model, regardless of whether or not energy injection is included. The CRs of SSC afterglow models are required to account for the bursts in the 2FLGC that cannot be explained in the synchrotron scenario, together with the study supplied by \cite{2019ApJ...883..134T} (e.g., those with photon energies above 10 GeV).

The afterglow of a GRB and the values of its spectral and temporal indices could be well explained by the standard synchrotron forward-shock model with the appropriate values of variation of the microphysical parameters, which is valid up to the synchrotron limit.   In contrast to the synchrotron model, which predicts that the maximum photon energy emitted during the deceleration phase will be around a few GeV, we argue that the SSC afterglow model is more suited to understanding this sample of bursts. Neutrino non-coincidences with GRBs have been detected by the IceCube Team \citep{2012Natur.484..351A, 2016ApJ...824..115A, 2015ApJ...805L...5A}, ruling out photo-hadronic interactions as a possible explanation. Although the CRs of the synchrotron afterglow model might, in some circumstances, characterize this sample of 29 bursts, the SSC process may be more appropriate.

As a particular case, we have applied the SSC scenario of the microphysical parameter variations evolving in a constant-density medium to describe the latest \textit{Fermi}-LAT observations of GRB 160509A. We have shown that these observations cannot be described with synchrotron flux but with SSC emission from the forward shock, where it evolves in a uniform-density environment.

We have applied the CRs of the SSC and synchrotron afterglow models in each cooling condition to the entire sample of GRBs reported in the 2FLGC using MCMC simulations. We have also applied this procedure to a selection of 4 bursts (GRB 180720B, 190114C, 190829A, and 221009A) that presented photons with energies above 100 GeV, which the MAGIC, H.E.S.S., and LHAASO gamma-ray observatories detected. From this analysis, we have concluded that variations of the spectral (temporal) index keeping the temporal (spectral) index constant are associated with the evolution of microphysical parameters.

\section*{Acknowledgements}

We appreciate the useful discussions with Tanmoy Laskar, Rodolofo Barniol-Duran, and Peter Veres.    NF is grateful to UNAM-DGAPA-PAPIIT for the funding provided by grant IN106521.

\section*{Data Availability}

No new data were generated or analysed in support of this research.

\bibliographystyle{mnras}
\bibliography{Bib_GRB190829A} 

\clearpage
\newpage
\begin{table*}

\centering \renewcommand{\arraystretch}{1.85}\addtolength{\tabcolsep}{1.5pt}
\caption{CRs of the synchrotron afterglow scenario with variation of microphysical parameters in a stratified environment.}
\label{Table1}

\begin{tabular}{c c c  c}
 \hline \hline
&\hspace{0.1cm}     &\hspace{0.1cm}   Synchrotron &\hspace{0.1cm}   Synchrotron \\ 
&\hspace{0.1cm}     &\hspace{0.1cm} $1<p<2$   &\hspace{0.1cm}   $2<p$\\ 
 
                     & \hspace{0.1cm}  $\beta $            &   \hspace{0.1cm}  $\alpha $  &  \hspace{0.1cm}  $\alpha $\\  \hline \hline
Fast cooling $(\nu^{\rm syn}_{\rm c} < \nu^{\rm syn}_{\rm m})$ \\ \hline

$\nu^{\rm syn}_{\rm c} < \nu < \nu^{\rm syn}_{\rm m} $   	                & \hspace{0.1cm} $\frac{1}{2} $    &\hspace{0.1cm} $\frac{(1-b)\beta}{2} $	                    &\hspace{0.1cm} $\frac{(1-b)\beta}{2} $\\ 	
$\nu^{\rm syn}_{\rm m} < \nu $   	                                 & \hspace{0.1cm} $\frac{p}{2} $     &\hspace{0.1cm} $\frac{5-2a(k-4)+3\beta-k(\beta+1)}{2(4-k)} $	                    &\hspace{0.1cm} $\frac{3\beta-1+b(\beta-1)+2a(2\beta-1)}{2} $\\ \hline
Slow cooling $(\nu^{\rm syn}_{\rm m} < \nu^{\rm syn}_{\rm c})$ \\\hline 	

$ \nu^{\rm syn}_{\rm m} < \nu < \nu^{\rm syn}_{\rm c} $   	                & \hspace{0.1cm} $\frac{p-1}{2} $  &\hspace{0.1cm} $\frac{9-(4a+3b)(k-4)+2\beta(3-k)}{4(4-k)} $	                    &\hspace{0.1cm} $\frac{12\beta+k(1-3\beta)+4a\beta(4-k)+b(4-k)(\beta+1)}{2(4-k)} $\\ 	
$\nu^{\rm syn}_{\rm c} < \nu $   	                                 & \hspace{0.1cm} $\frac{p}{2} $     &\hspace{0.1cm} $\frac{5-2a(k-4)+3\beta-k(\beta+1)}{2(4-k)} $	                    &\hspace{0.1cm} $\frac{3\beta-1+b(\beta-1)+2a(2\beta-1)}{2} $\\ \hline

%
%
\end{tabular}
\end{table*}

\begin{table*}

\centering \renewcommand{\arraystretch}{1.85}\addtolength{\tabcolsep}{1.5pt}
\caption{CRs of the SSC afterglow scenario with variation of microphysical parameters in a stratified environment.}
\label{Table2}

\begin{tabular}{c c c  c}
 \hline \hline
&\hspace{0.1cm}      & \hspace{0.1cm}   SSC   & \hspace{0.1cm}  SSC \\ 
&\hspace{0.1cm}       & \hspace{0.1cm}    $1<p<2$   & \hspace{0.1cm}  $2<p$    \\ 
 
                     & \hspace{0.1cm}  $\beta $            & \hspace{0.1cm}  $\alpha $ & \hspace{0.1cm} $\alpha $ \\  \hline \hline
Fast cooling $(\nu^{\rm ssc}_{\rm c} < \nu^{\rm ssc}_{\rm m})$ \\ \hline

$\nu^{\rm ssc}_{\rm c} < \nu < \nu^{\rm ssc}_{\rm m} $   	                & \hspace{0.1cm} $\frac{1}{2} $    &\hspace{0.1cm} $\frac{[k-5b(4-k)-2]\beta}{2(4-k)} $ &\hspace{0.1cm} $\frac{[k-5b(4-k)-2]\beta}{2(4-k)}$ \\ 	
$\nu^{\rm ssc}_{\rm m} < \nu $   	                                 & \hspace{0.1cm} $\frac{p}{2} $     &\hspace{0.1cm} $\frac{8-4a(k-4)-k(\beta+1)+b(k-4)(\beta+1)}{2(4-k)} $  &\hspace{0.1cm} $\frac{2(9\beta-5)+k(3-5\beta)+b(4-k)(\beta-3)+4a(4-k)(2\beta-1)}{2(4-k)}$\\ \hline
Slow cooling $(\nu^{\rm ssc}_{\rm m} < \nu^{\rm ssc}_{\rm c})$ \\\hline 	

$ \nu^{\rm ssc}_{\rm m} < \nu < \nu^{\rm ssc}_{\rm c} $   	                & \hspace{0.1cm} $\frac{p-1}{2} $  &\hspace{0.1cm} $\frac{7-4a(k-4)+k(1-\beta)+b(k-4)(\beta-2)}{2(4-k)} $ &\hspace{0.1cm} $\frac{2(9\beta-1)+k(3-5\beta)+8a\beta(4-k)+b(4-k)(\beta+1)}{2(4-k)}$ \\ 	
$\nu^{\rm ssc}_{\rm c} < \nu $   	                                 & \hspace{0.1cm} $\frac{p}{2} $     &\hspace{0.1cm} $\frac{8-4a(k-4)-k(\beta+1)+b(k-4)(\beta+1)}{2(4-k)} $ &\hspace{0.1cm} $\frac{2(9\beta-5)+k(3-5\beta)+b(4-k)(\beta-3)+4a(4-k)(2\beta-1)}{2(4-k)}$\\ \hline

%
%
\end{tabular}
\end{table*}


\begin{table*}
    \centering\renewcommand{\arraystretch}{1.4}
    \caption{Results from MCMC simulations of SSC and synchrotron afterglow model evolving in a stratified circumburst medium.}
    \label{Table3}
    \begin{tabular}{c c c c c c c}
     \hline
     \hline
                  &      & Synchrotron  &  & & SSC &  \\
      $\nu$ Range & $k$  & a  & b &  & a & b \\
     \cline{1-4} \cline{6-7}
    ${\rm \nu_m^{\rm j} < \nu < \nu_{c}^{\rm j}}$ & $0.0$ &  $-1.50\substack{+0.02 \\ -0.02}$ & $2.77\substack{+0.06 \\ -0.06}$ & & $-1.00\substack{+0.01 \\ -0.01}$ & $3.27\substack{+0.06 \\ -0.06}$ \\\cline{2-4} \cline{6-7}
     & 0.5 &  $-1.46\substack{+0.02 \\ -0.02}$ & $2.64\substack{+0.06 \\ -0.06}$ & & $-0.94\substack{+0.01 \\ -0.01}$ & $2.91\substack{+0.06 \\ -0.06}$ \\\cline{2-4} \cline{6-7}
     & 1.0 &  $-1.42\substack{+0.02 \\ -0.02}$ & $2.45\substack{+0.06 \\ -0.06}$ & & $-0.87\substack{+0.01 \\ -0.01}$ & $2.44\substack{+0.06 \\ -0.06}$ \\\cline{2-4} \cline{6-7}
     & 1.5 &  $-1.35\substack{+0.02 \\ -0.02}$ & $2.18\substack{+0.06 \\ -0.06}$ & & $-0.77\substack{+0.01 \\ -0.01}$ & $1.77\substack{+0.06 \\ -0.06}$ \\\cline{2-4} \cline{6-7}
     & 2.0 &  $-1.25\substack{+0.02 \\ -0.02}$ & $1.78\substack{+0.06 \\ -0.06}$ & & $-0.62\substack{+0.01 \\ -0.01}$ & $0.77\substack{+0.06 \\ -0.06}$ \\\cline{2-4} \cline{6-7}
     & 2.5 &  $-1.08\substack{+0.02 \\ -0.02}$ & $1.11\substack{+0.06 \\ -0.06}$ & & $-0.37\substack{+0.01 \\ -0.01}$ & $-0.89\substack{+0.06 \\ -0.06}$ \\\cline{2-4} \cline{6-7}

    \hline
    \hline
     ${\rm \nu_c^{\rm j} < \nu < \nu_{m}^{\rm j}}$ & 0.0 & $0.03\substack{+10.17 \\ -10.25}$ & $-3.98\substack{+0.03 \\ -0.03}$ & & $-0.01\substack{+3.41 \\ -3.41}$ & $-1.08\substack{+0.01 \\ -0.01}$ \\\cline{2-4} \cline{6-7}
     & 0.5  & $0.15\substack{+10.15 \\ -10.26}$ & $-3.89\substack{+0.03 \\ -0.03}$ & & $-0.01\substack{+3.42 \\ -3.41}$ & $-1.06\substack{+0.01 \\ -0.01}$ \\\cline{2-4} \cline{6-7}
     & 1.0  & $0.04\substack{+10.21 \\ -10.24}$ & $-3.89\substack{+0.03 \\ -0.03}$ & & $0.00\substack{+3.40 \\ -3.40}$ & $-1.05\substack{+0.01 \\ -0.01}$ \\\cline{2-4} \cline{6-7}
     &  1.5 & $0.17\substack{+10.11 \\ -10.29}$ & $-3.89\substack{+0.03 \\ -0.03}$ & & $-0.02\substack{+3.41 \\ -3.39}$ & $-1.02\substack{+0.01 \\ -0.01}$ \\\cline{2-4} \cline{6-7}
     & 2.0  & $0.10\substack{+10.23 \\ -10.15}$ & $-3.89\substack{+0.03 \\ -0.03}$ & & $-0.01\substack{+3.42 \\ -3.39}$ & $-0.98\substack{+0.01 \\ -0.01}$ \\\cline{2-4} \cline{6-7}
     & 2.5  & $0.03\substack{+10.25 \\ -10.19}$ & $-3.89\substack{+0.03 \\ -0.03}$ & & $0.02\substack{+3.39 \\ -3.41}$ & $-0.91\substack{+0.01 \\ -0.01}$ \\\cline{2-4} \cline{6-7}
     
     \hline
     \hline
     ${\rm max\{\nu_m^{\rm j}, \nu_c^{\rm j}\} < \nu}$ & 0.0  & $0.34\substack{+0.02 \\ -0.02}$ & $-4.74\substack{+0.10 \\ -0.11}$ & & $-0.46\substack{+0.01 \\ -0.01}$ & $-1.28\substack{+0.02 \\ -0.02}$ \\\cline{2-4} \cline{6-7}
     & 0.5  & $0.34\substack{+0.01 \\ -0.01}$ & $-4.73\substack{+0.09 \\ -0.09}$ & & $-0.45\substack{+0.01 \\ -0.01}$ & $-1.26\substack{+0.02 \\ -0.02}$ \\\cline{2-4} \cline{6-7}
     & 1.0  & $0.34\substack{+0.01 \\ -0.01}$ & $-4.71\substack{+0.09 \\ -0.09}$ & & $-0.44\substack{+0.01 \\ -0.01}$ & $-1.23\substack{+0.02 \\ -0.02}$ \\\cline{2-4} \cline{6-7}
     &  1.5 & $0.33\substack{+0.01 \\ -0.01}$ & $-4.69\substack{+0.09 \\ -0.09}$ & & $-0.42\substack{+0.01 \\ -0.01}$ & $-1.18\substack{+0.02 \\ -0.02}$ \\\cline{2-4} \cline{6-7}
     & 2.0  & $0.33\substack{+0.01 \\ -0.01}$ & $-4.67\substack{+0.09 \\ -0.09}$ & & $-0.40\substack{+0.01 \\ -0.01}$ & $-1.12\substack{+0.02 \\ -0.02}$ \\\cline{2-4} \cline{6-7}
     & 2.5  & $0.32\substack{+0.01 \\ -0.01}$ & $-4.62\substack{+0.09 \\ -0.09}$ & & $-0.36\substack{+0.01 \\ -0.01}$ & $-1.02\substack{+0.02 \\ -0.02}$ \\\cline{2-4} \cline{6-7}
     \hline
     \hline

    \end{tabular}

\end{table*}

\begin{table*}
    \centering\renewcommand{\arraystretch}{1.4}
    \caption{Number and percentage of bursts satisfying each CR, summarized for the synchrotron forward-shock model with $1<p<2$ and $2<p$. We consider the best-fit values of $a$ and $b$ obtained with our MCMC and reported in Table \ref{Table3}.}
    \label{Table4}
    \begin{tabular}{c c c c c c}
     \hline
     \hline
    
      $\nu$ Range & $k$  & CR: 1 $<$ p $<$ 2  & CR:  2 $<$ p & GRBs Satisfying Relation & Proportion Satisfying Relation \\
                  &      &                    &              &           &   (\%)                   \\
     \hline
     
    ${\rm \nu_m^{\rm syn} < \nu < \nu_{c}^{\rm syn}}$ & $0.0$ &  $\frac{16a+3(3+4b+2\beta)}{16}$ & $\frac{b+(3+4a+b)\beta}{2}$ & 23 & 27.06 \\\cline{2-6}
     & 0.5 &   $\frac{18+28a+21b+10\beta}{28}$  & $\frac{1+7b+7(3+4a+b)\beta}{14}$  & 22 & 25.88 \\\cline{2-6}
     & 1.0&   $\frac{9+12a+9b+4\beta}{12}$ & $\frac{1+3b+3(3+4a+b)\beta}{6}$  & 23 & 27.06 \\\cline{2-6}
     & 1.5 &   $\frac{18+20a+15b+6\beta}{20}$  & $\frac{3+5b+5(3+4a+b)\beta}{10}$  & 23 & 27.06 \\\cline{2-6}
     & 2.0&   $\frac{9+8a+6b+2\beta}{8}$ & $\frac{1+b+(3+4a+b)\beta}{2}$  & 23 & 27.06 \\\cline{2-6}
     & 2.5 &   $\frac{18+12a+9b+2\beta}{12}$  & $\frac{5+3b+3(3+4a+b)\beta}{6}$  & 22 & 25.88 \\\cline{2-6}

    \hline
    \hline
     ${\rm \nu_c^{\rm syn} < \nu < \nu_{m}^{\rm syn}}$ & 0.0 & $\frac{(1-b)\beta}{2}$ & $\frac{(1-b)\beta}{2}$ & 1 & 1.18 \\\cline{2-6}
     & 0.5  & $\frac{(1-b)\beta}{2}$  & $\frac{(1-b)\beta}{2}$ & 1 & 1.18 \\\cline{2-6}
     & 1.0  & $\frac{(1-b)\beta}{2}$  & $\frac{(1-b)\beta}{2}$ & 1 & 1.18 \\\cline{2-6}
     &  1.5  & $\frac{(1-b)\beta}{2}$ & $\frac{(1-b)\beta}{2}$ & 1 & 1.18 \\\cline{2-6}
     & 2.0&   $\frac{(1-b)\beta}{2}$ & $\frac{(1-b)\beta}{2}$  & 1 & 1.18 \\\cline{2-6}
     & 2.5 &   $\frac{(1-b)\beta}{2}$  & $\frac{(1-b)\beta}{2}$  & 1 & 1.18 \\\cline{2-6}
     
     \hline
     \hline
     ${\rm max\{\nu_m^{\rm syn}, \nu_c^{\rm syn}\} < \nu}$ & 0.0  & $\frac{5+8a+3\beta}{8}$ & $-\frac{1+2a+b-(3+4a+b)\beta}{2}$ & 23 & 27.06 \\\cline{2-6}
     & 0.5  & $\frac{9+14a+5\beta}{14}$  & $-\frac{1+2a+b-(3+4a+b)\beta}{2}$ & 23 & 27.06 \\\cline{2-6}
     & 1.0  & $\frac{2+3a+\beta}{3}$  & $-\frac{1+2a+b-(3+4a+b)\beta}{2}$ & 23 & 27.06 \\\cline{2-6}
     &  1.5 & $\frac{7+10a+3\beta}{10}$ & $-\frac{1+2a+b-(3+4a+b)\beta}{2}$ & 23 & 27.06 \\\cline{2-6}
     & 2.0 &   $\frac{3+4a+\beta}{4}$  & $-\frac{1+2a+b-(3+4a+b)\beta}{2}$  & 22 & 25.88 \\\cline{2-6}
     & 2.5&   $\frac{5+6a+\beta}{6}$ & $-\frac{1+2a+b-(3+4a+b)\beta}{2}$  & 22 & 25.88 \\
     \hline
     \hline

    \end{tabular}

\end{table*}

\begin{table*}
    \centering\renewcommand{\arraystretch}{1.8}
    \caption{Number and percentage of bursts satisfying each CR, summarized for the SSC forward-shock model with $1<p<2$ and $2<p$. We consider the best-fit values of $a$ and $b$ obtained with our MCMC and reported in Table \ref{Table3}.}
    \label{Table5}
    \begin{tabular}{c c c c c c}
     \hline
     \hline
    
      $\nu$ Range & $k$  & CR: 1 $<$ p $<$ 2  & CR:  2 $<$ p & GRBs Satisfying Relation & Proportion Satisfying Relation \\
                  &      &                    &              &           &   (\%)                   \\
     \hline
     
    ${\rm \nu_m^{\rm ssc} < \nu < \nu_{c}^{\rm ssc}}$ & $0.0$ &  $\frac{7+16a+4b(2-\beta)}{8}$ & $-\frac{1-(9+16a)\beta-2b(1+\beta)}{4}$ & 23 & 27.06 \\\cline{2-6}
     & 0.5 &   $\frac{15+28a+7b(2-\beta)-\beta}{14}$  & $-\frac{1-(31+56a)\beta-7b(1+\beta)}{14}$  & 22 & 25.88 \\\cline{2-6}
     & 1.0&   $\frac{4(2+3a)+3b(2-\beta)-\beta}{6}$ & $\frac{1+(13+24a)\beta+3b(1+\beta)}{6}$  & 22 & 25.88 \\\cline{2-6}
     & 1.5 &   $\frac{17+20a+5b(2-\beta)-3\beta}{10}$  & $\frac{5+(21+40a)\beta+5b(1+\beta)}{10}$  & 22 & 25.88 \\\cline{2-6}
     & 2.0&   $\frac{9+8a+2b(2-\beta)-2\beta}{4}$ & $\frac{2+b+(4+8a+b)\beta}{2}$  & 22 & 25.88 \\\cline{2-6}
     & 2.5 &   $\frac{19+12a+3b(2-\beta)-5\beta}{6}$  & $\frac{11+(11+24a)\beta+3b(1+\beta)}{6}$  & 22 & 25.88 \\\cline{2-6}

    \hline
    \hline
     ${\rm \nu_c^{\rm ssc} < \nu < \nu_{m}^{\rm ssc}}$ & 0.0 & $-\frac{(1+10b)\beta}{4}$ & $-\frac{(1+10b)\beta}{4}$ & 1 & 1.18 \\\cline{2-6}
     & 0.5  & $-\frac{(3+35b)\beta}{14}$  & $-\frac{(3+35b)\beta}{14}$ & 1 & 1.18 \\\cline{2-6}
     & 1.0  & $-\frac{(1+15b)\beta}{6}$  & $-\frac{(1+15b)\beta}{6}$ & 1 & 1.18 \\\cline{2-6}
     &  1.5  & $-\frac{(1+25b)\beta}{10}$ & $-\frac{(1+25b)\beta}{10}$ & 1 & 1.18 \\\cline{2-6}
     & 2.0&   $-\frac{5b\beta}{2}$ & $-\frac{5b\beta}{2}$  & 1 & 1.18 \\\cline{2-6}
     & 2.5 &   $\frac{(1-15b)\beta}{6}$  & $\frac{(1-15b)\beta}{6}$  & 1 & 1.18 \\\cline{2-6}
     
     \hline
     \hline
     ${\rm max\{\nu_m^{\rm ssc}, \nu_c^{\rm ssc}\} < \nu}$ & 0.0  & $\frac{2(1+2a)-b(1+\beta)}{2}$ & $-\frac{5+2b(3-\beta)-9\beta+8a(1-2\beta)}{4}$ & 24 & 28.24 \\\cline{2-6}
     & 0.5  & $\frac{15+28a-7b(1+\beta)-\beta}{14}$  & $-\frac{17+7b(3-\beta)-31\beta+28a(1-2\beta)}{14}$ & 24 & 28.24 \\\cline{2-6}
     & 1.0  & $\frac{7+12a-3b(1+\beta)-\beta}{6}$  & $-\frac{7+3b(3-\beta)-13\beta+12a(1-2\beta)}{6}$ & 23 & 27.06 \\\cline{2-6}
     &  1.5 & $\frac{13+20a-5b(1+\beta)-3\beta}{10}$ & $-\frac{11+5b(3-\beta)-21\beta+20a(1-2\beta)}{10}$ & 23 & 27.06 \\\cline{2-6}
     & 2.0 &   $\frac{3+4a-b(1+\beta)-\beta}{2}$  & $-\frac{2+4a+3b-(4+8a+b)\beta}{2}$  & 21 & 24.71 \\\cline{2-6}
     & 2.5&   $\frac{11+12a-3b(1+\beta)-5\beta}{6}$ & $-\frac{5+3b(3-\beta)-11\beta+12a(1-2\beta)}{6}$  & 21 & 24.71 \\
     \hline
     \hline

    \end{tabular}

\end{table*}

\begin{table*}
\centering \renewcommand{\arraystretch}{1.85}\addtolength{\tabcolsep}{1.5pt}
\caption{Evolution of the synchrotron light curves ($F_\nu\propto t^{-\alpha}\nu^{-\beta}$) from an outflow decelerated in a stratified environment with microphysical parameter evolution}\label{Table:sync}
\begin{tabular}{cccccc}
\hline
\hline
 &  & \multicolumn{2}{c}{$k=0$} & \multicolumn{2}{c}{$k=2$} \\
 &  & $1<p<2$ & $2<p$ & $1<p<2$ & $2<p$ \\
 & $\beta$ & $\alpha$ & $\alpha$ & $\alpha$ & $\alpha$ \\
 \hline
 \hline
Fast cooling $(\nu^{\rm syn}_{\rm c} < \nu^{\rm syn}_{\rm m})$ &  &  &  &  &  \\
\hline
$ \nu < \nu^{\rm syn}_{\rm c} $ & $-\frac13$ & $\frac{6b-1}{6} $ & $\frac{6b-1}{6} $ & $\frac{3b+2}{3}$ & $\frac{3b+2}{3}$ \\
$\nu^{\rm syn}_{\rm c} < \nu < \nu^{\rm syn}_{\rm m} $ & $\frac12$ & $\frac{1-b}{4} $ & $\frac{1-b}{4} $ & $\frac{1-b}{4} $ & $\frac{1-b}{4} $ \\
$\nu^{\rm syn}_{\rm m} < \nu $ & $\frac{p}{2}$ & $\frac{10+16a+3p}{16} $ & $-\frac{2(1+2a+b)-p(3+4a+b)}{4} $ & $\frac{6+8a+p}{8} $ & $-\frac{2(1+2a+b)-p(3+4a+b)}{4} $ \\
\hline
Slow cooling $(\nu^{\rm syn}_{\rm m} < \nu^{\rm syn}_{\rm c})$ &  &  &  &  &  \\
\hline
$\nu < \nu^{\rm syn}_{\rm m} $ & $-\frac{1}{3}$ & $-\frac{3(2+p)+16a-4b(3p-4)}{24(p-1)} $ & $-\frac{3-2(b-2a)}{6} $ & $-\frac{5(2-p)+8a-2b(3p-4)}{12(p-1)} $ & $-\frac{2a-b}{3}$ \\
$ \nu^{\rm syn}_{\rm m} < \nu < \nu^{\rm syn}_{\rm c} $ & $\frac{p-1}{2}$ & $\frac{3(2+p)+4(4a+3b)}{16} $ & $-\frac{3+4a-b-p(3+4a+b)}{4} $ & $\frac{8+p+2(4a+3b)}{8} $ & $-\frac{1+4a-b-p(3+4a+b)}{4} $ \\
$\nu^{\rm syn}_{\rm c} < \nu $ & $\frac{p}{2}$ & $\frac{10+16a+3p}{16} $ & $-\frac{2(1+2a+b)-p(3+4a+b)}{4} $ & $\frac{6+8a+p}{8} $ & $-\frac{2(1+2a+b)-p(3+4a+b)}{4} $ \\
\hline
\end{tabular}
\end{table*}

\begin{table*}
\centering \renewcommand{\arraystretch}{1.85}\addtolength{\tabcolsep}{1.5pt}
\caption{Evolution of the SSC light curves ($F_\nu\propto t^{-\alpha}\nu^{-\beta}$) from an outflow decelerated in a stratified environment with microphysical parameter evolution}\label{Table:ssc}
\begin{tabular}{cccccc}
\hline
\hline
 &  & \multicolumn{2}{c}{$k=0$} & \multicolumn{2}{c}{$k=2$} \\
 &  & $1<p<2$ & $2<p$ & $1<p<2$ & $2<p$ \\
 & $\beta$ & $\alpha$ & $\alpha$ & $\alpha$ & $\alpha$ \\
 \hline
 \hline
Fast cooling $(\nu^{\rm ssc}_{\rm c} < \nu^{\rm ssc}_{\rm m})$ &  &  &  &  &  \\
\hline
$\nu^{\rm ssc}_{\rm c} < \nu < \nu^{\rm ssc}_{\rm m} $ & $\frac12$ & $-\frac{10b+1}{8} $ & $-\frac{10b+1}{8} $ & $-\frac{5b}{4} $ & $-\frac{5b}{4} $ \\
$\nu^{\rm ssc}_{\rm m} < \nu $ & $\frac{p}{2}$ & $\frac{4(1+2a)-b(2+p)}{4} $ & $-\frac{2(5+8a+6b)-p(9+16a+2b)}{8}$ & $\frac{2(3+4a)-p-b(2+p)}{4} $ & $-\frac{2(2+4a+3b)-p(4+8a+b)}{4}$ \\
\hline
Slow cooling $(\nu^{\rm ssc}_{\rm m} < \nu^{\rm ssc}_{\rm c})$ &  &  &  &  &  \\
\hline
$ \nu^{\rm ssc}_{\rm m} < \nu < \nu^{\rm ssc}_{\rm c} $ & $\frac{p-1}{2}$ & $\frac{7+16a+b(10-2p)}{8} $ & $-\frac{11+16a-2b-p(9+16a+2b)}{8}$ & $\frac{2(5+4a)-p+b(5-p)}{4} $ & $-\frac{8a-b-p(4+8a+b)}{4}$ \\
$\nu^{\rm ssc}_{\rm c} < \nu $ & $\frac{p}{2}$ & $\frac{4(1+2a)-b(2+p)}{4} $ & $-\frac{2(5+8a+6b)-p(9+16a+2b)}{8}$ & $\frac{2(3+4a)-p-b(2+p)}{4} $ & $-\frac{2(2+4a+3b)-p(4+8a+b)}{4}$ \\
\hline
\end{tabular}
\end{table*}

\begin{table}
\centering \renewcommand{\arraystretch}{1.5}\addtolength{\tabcolsep}{2pt}

\caption{Sample of 8 bursts from 2FLGC. Temporal and spectral PL indices are taken from \citet{2019ApJ...878...52A} with $\beta_{\rm L}=\Gamma_{\rm L}-1$.}
\label{tab:sample}
\begin{tabular}{lcc}
\hline
GRB & $\alpha_{\rm L} \pm \delta_{\alpha_{\rm L}}$ & $\beta_{\rm L} \pm \delta_{\beta_{\rm L}}$ \\
\hline
090902B	&	$1.63	\pm	0.08$	&	$0.92	\pm	0.06$		\\
090926A	&	$1.82	\pm	0.08$	&	$0.86	\pm	0.07$			\\
100116A	&	$2.70	\pm	0.20$	&	$0.60	\pm	0.20$			\\
100414A	&	$1.30	\pm	0.10$	&	$0.80	\pm	0.10$			\\
130327B	&	$1.60	\pm	0.20$	&	$0.80	\pm	0.10$			\\
141207A	&	$1.88	\pm	0.03$	&	$0.80	\pm	0.30$			\\
160625B	&	$2.20	\pm	0.30$	&	$0.80	\pm	0.30$			\\
180526A	&	$1.30	\pm	0.70$	&	$0.80	\pm	0.30$			\\

\hline
\end{tabular}
\end{table}

\begin{table}
\centering \renewcommand{\arraystretch}{1.5}\addtolength{\tabcolsep}{2pt}

\caption{Sample of 4 bursts detected with VHE photons. Temporal and spectral PL indices are taken from \citet{HESS_2021Sci...372.1081H, MAGIC_2019Natur.575..459M, MAGIC_2019Natur.575..455M, 2019Natur.575..464A, 2023Sci...380.1390L} with $\beta_{\rm VHE}=\Gamma_{\rm VHE}-1$.}
\label{tab:sample_vhe}
\begin{tabular}{lcccc}
\hline
GRB & $\alpha_{\rm L} \pm \delta_{\alpha_{\rm L}}$ & $\beta_{\rm L} \pm \delta_{\beta_{\rm L}}$ & $\alpha_{\rm VHE} \pm \delta_{\alpha_{\rm VHE}}$ & $\beta_{\rm VHE} \pm \delta_{\beta_{\rm VHE}}$ \\
\hline
180720B	&	$1.90	\pm	0.10$	&	$1.15	\pm	0.10$	&	$-$	&	$-$	\\
190114C	&	$1.10	\pm	0.15$	&	$1.06	\pm	0.30$	&	$1.60	\pm	0.07$	&	$1.22	\pm	0.23$		\\
190829A &	$-$	&	$-$	&	$1.09	\pm	0.05$	&	$1.06	\pm	0.10$ $^{*}$		\\
        &	$-$	&	$-$	&	$1.09	\pm	0.05$	&	$0.86	\pm	0.26$ $^{**}$		\\
221009A	&	$1.32	\pm	0.05$	&	$0.87	\pm	0.04$	&	$1.115	\pm	0.012$	&	$1.320	\pm	0.015$		\\

\hline
\end{tabular}
\\$^*$ These values correspond to the first night of observations.
\\$^{**}$ These values correspond to the second night of observations.
\end{table}


\clearpage

\bsp	
\label{lastpage}
\end{document}